\documentclass[prd,superscriptaddress,amsfonts,amssymb,amsmath,showpacs,twocolumn,nofootinbib]{revtex4-2}
\usepackage{eurosym}
\usepackage{amsmath}
\usepackage{bm}
\usepackage{amsfonts}
\usepackage{amssymb}
\usepackage{graphicx}
\usepackage{mathtools}
\usepackage{color}
\usepackage{booktabs}
\usepackage{physics}

\def\be{\begin{equation}}
\def\ee{\end{equation}}
\def\bea{\begin{eqnarray}}
\def\eea{\end{eqnarray}}

\begin{document}
\title{Quasinormal Modes of Black holes in the generalized Schwarzschild-de Sitter solution of $f(R)$ gravity}

\author{Dhruba Jyoti Gogoi}
\email{moloydhruba@yahoo.in}
\affiliation{Department of Physics, Madhabdev University, Narayanpur, Lakhimpur 784164, Assam, India}
\affiliation{Research Center of Astrophysics and Cosmology, Khazar University, Baku, AZ1096, 41 Mehseti Street, Azerbaijan}

\author{Himanshu Chaudhary}
\email{himanshu.chaudhary@ubbcluj.ro,\\
himanshuch1729@gmail.com}
\affiliation{Faculty of Physics, Babeș-Bolyai University, Kogălniceanu Street, Cluj-Napoca, 400084, Romania,}

\author{Tiberiu Harko}
\email{tiberiu.harko@aira.astro.ro}
\affiliation{Faculty of Physics, Babeș-Bolyai University, Kogălniceanu Street, Cluj-Napoca, 400084, Romania,}
\affiliation{Astronomical Observatory, 19 Cire\c silor Street, Cluj-Napoca 400487, Romania.}

\begin{abstract}
We analyze scalar, vector, and {axial} tensor perturbations in a Schwarzschild-type static spherically symmetric solution of modified $f(R)$ gravity. This solution differs from the Schwarzschild metric of general relativity through the presence of a linearly growing term and a logarithmic correction, and it depends on four integration constants. Using the higher-order Wentzel-Kramers-Brillouin (WKB) approximation and an analytical method, we calculate the quasinormal modes (QNMs) for all three types of perturbations. Our results show that, for appropriately chosen parameter values, the QNMs closely match those of the Schwarzschild black hole. In particular, the model parameter $A$ produces a nonlinear increase in both the oscillation frequency and damping rate of ringdown gravitational waves. A similar effect arises for the parameter $\eta$ when $A$ is small, while for larger $A$ values the effects are opposite, where the scalar perturbations display a distinct deviation compared to the vector and {axial} tensor cases. Contour plots of QNMs in the parameter space support these findings. Finally, time-domain profiles are obtained to illustrate the dynamical evolution of the gravitational wave signal.

\end{abstract}


\maketitle

\textcolor{blue}{\tableofcontents}

\section{Introduction}

The discovery of the recent acceleration of the Universe \cite{O1,O2,O3,O4}, as well as the convincing observational evidence for the existence of dark matter, raise the fundamental question of the validity of Einstein's general relativity at astrophysical and cosmological scales. Still within the theoretical framework of general relativity, a simple and yet powerful cosmological model, the concordance or standard $\Lambda$CDM model was proposed \cite{L1,L2,L3}, which explains the late cosmological acceleration by the presence of a cosmological constant $\Lambda$. $\Lambda$ is assumed to represent a homogeneous form of energy, independent of the cosmological expansion, and  usually considered as a particular case of the dark energy, or as a vacuum energy. For reviews of the $\Lambda$CDM model see \cite{L4,L5,L6,L7}. However, the $\Lambda$CDM model faces a number of significant challenges, like the absence of a theoretical understanding of $\Lambda$, the Hubble $H_0$ tension, the growth $\sigma _8$ tension, the lensing amplitude $A_L$ anomaly etc. \cite{diV,Pou,Scol}.  Thus, the search for alternative explanations for observational data remains a central theoretical problem in gravitational physics, and cosmology.

Perhaps the simplest possibility for explaining the observational data is to assume that on  large scales, going beyond the scale of the Solar System,  the Einstein gravity theory breaks down, and therefore one needs to adopt a more general approach for describing the gravitational field. Many modified theories of gravity have been proposed, and for detailed reviews of these theories see \cite{M1,M2,M3, M4}. 

One of the first modified gravity theories that have been proposed in the literature is the $f(R)$ gravity theory \cite{Bu}. While general relativity is based on the Hilbert-Einstein action $S_{HE}=\int{\left(R+L_m\right)\sqrt{-g}d^4x}$, where $R$ is the Ricci scalar, and $L_m$ is the matter Lagrangian, in the $f(R)$ theory the standard Hilbert-Einstein action is generalized by assuming for the gravitational action an arbitrary function of the Ricci scalar $R$. Hence the action of the gravitational sector of the theory is given by $S=\int{f(R)\sqrt{-g}d^4x}$. The $f(R)$ theory was first proposed in \cite{Bu}, and its first implications were later considered in \cite{Ke,Ke1,Ba}. The cosmological applications of the theory was investigated in an early study \cite{Kl}. The only important restriction assumed on the function $f(R)$ is that it must  be analytic, and therefore it must admit a Taylor series expansion in any point $R$. $f(R)$ theory has a very important theoretical potential for the understanding of the cosmological observational data. It can explain the  cosmic acceleration through the geometry of space-time itself, by adopting specific functional forms for $f(R)$, without the need  of introducing a dark energy component. However, $f(R)$ theories require self tuning to satisfy both cosmological observations and the local gravity tests, as shown in the early studies  \cite{fr1,fr2,fr3,fr4,fr5, fr6}. For the cosmological and astrophysical applications of the $f(R)$ gravity theories see \cite{fr7,fr7a, fr8,fr9,fr10,fr11,fr11a, fr12,fr13,fr14, fr14a, fr15,fr16, fr16a, fr17,fr18,fr19,fr20,fr21,fr21a, fr22,fr23,fr24,fr25}.

The first static spherically symmetric solution of the Einstein field equations was obtained by Schwarzschild in 1916 \cite{bh1}. The solution describes the gravitational field of a  point-like, static, spherically symmetric  mass distribution. The Schwarzschild solution introduced a new theoretical concepts, related to a specific type of astrophysical objects, called black holes, having intriguing and baffling astrophysical properies. It took 60 years until the external solution of the Einstein equations for a rotating object was obtained by Kerr \cite{bh2}. From a theoretical perspective black holes are objects with a very simple structure, with two essential characteristics, a central singularity, and an event horizon \cite{chandrasekhar}. The existence of a limiting mass for neutron stars of the order of 3.2$M_{\odot}$ \cite{bh3} is an important indicator for the observational identification of black holes. 

A decisive test for the existence of black holes was presented by the Event Horizon Telescope collaboration, who obtained the first images of the supermassive black hole M87* \cite{bh4,bh5}. The observations seem to confirm that Sgr A* black hole is of Kerr type. But these observations still cannot exclude the possibility of the existence of at least some small departures with regard to the predictions of general relativity. Some fundamental, and yet unsolved  problems are also raised by the Schwarzschild and the Kerr solutions, related to the presence of singularities in black hole solutions \cite{bh6,bh7,bh8,Q5,Q6,Q4, Q10}.

The study of the black hole and static spherically symmetric solutions in $f(R)$ gravity proved to be an active and interesting field of research \cite{s0, s0a,s0b, s1,s2,s3,s3a, s4,s5,s6,s7,s8,s9,s10}.  In particular, in \cite{s0} an exact solution of the field equations of the $f(R)$ gravity was obtained, after reducing the system of the modified Einstein field equations to a single, third order differential equation.  A generalized Schwarzschild-de Sitter type solution of the $f(R)$ gravity vacuum field equations in spherical symmetry can thus be obtained, which adds to the standard form  a term linearly increasing with the radial coordinate, and a logarithmic term. The solution depends on four integration constants. The Schwarzschild de Sitter solution of standard general relativity  is reobtained by choosing the zero value for one integration constant, and by appropriately fixing the values of the other constants. This choice of forms of the integration constants guarantees that the model passes the Solar System tests. On the other hand,  at large distances from the central object, or for strong gravitational fields, the generalized Schwarzschild-de Sitter geometry of the $f(R)$ gravity may introduce some significant deviations as compared with the predictions of standard general relativity. 

The basic physical properties of accretion disks in the generalized Schwarzschild-de Sitter static and spherically symmetric geometry  of the vacuum $f(R)$ gravity were considered in \cite{s0a}, where the energy flux and the emission spectrum from the accretion disk were computed. A comparison with the the general relativistic case was also performed. Specific signatures were identified  in the electromagnetic spectrum, which indicates the possibility of the direct test of the $f(R)$ gravity model by using astrophysical observations of the emission spectra from accretion disks.

Black holes do not exist as isolated astrophysical objects. They are interacting with their astrophysical environments, by forming accretion disks, and they are strongly influenced by the presence of magnetic fields, and of the dark matter halos. The understanding of the observable quantities related to black hole physics is essential, since, among other observations,  precision gravitational wave astronomy has led to a large number of observational data. 

A key observational signature of black holes is provided by the spectrum of their Quasinormal Modes (QNMs) \cite{Qn1,Qn2,Qn3, Qn4, Qn5}, which describe the ringdown phase of a merger, and characterizes the late-time behavior of perturbations.

Quasi-normal modes can be interpreted as giving the characteristic sound of black holes under linear perturbations \cite{Qn2}. Moreover, since their spectra includes the fundamental physical parameters of the black hole, and they are fully determined by the background geometry, they can also be considered  as representing  the fingerprints of the black
holes. QNMs  provide  an important tool for identifying specific signatures of the space-time geometry. They play an important role in the gravitational wave physics, since the ring-down phase dominated by QNMs can give essential information for the identification of the type and of the parameters of the black holes. Moreover, QNMs can be used  for testing  the basic principles of general relativity, and alternative gravitational theories \cite{Te1,Te2,Te3}. The high-frequency limits of the QNMs spectra are closely related to the properties of the unstable photon orbits, indicating the existence of a close relationship between geometric optics and wave behavior \cite{Te4,Te5,Te6}. Moreover, QNMs play an important role in string theory, and especially in holographic duality, since they represent  the poles of the retarded Green functions, and contain information about quantum chaos and the thermalization rates \cite{Te7,Te8,Te9}. Thus, the study of QNMs can open a new window on the structure and dynamical properties of the black holes \cite{n0, Press, n1, mashhoon_1, mashhoon_2, mashhoon_3, Schutz, Will_wkb, Iyer, gundlach, n2,n3,n4,n5,n6,n7,n8,n9,n10,n11,n12,n13, n14,n15,n16,n17,n18,n19,n20,n21,lopez2020,n21,Hu, n22,n23,n24,n25,n26,n27,n28,n29, n30,n31,n32,n33,n34,n35,n36,n37,n38,n39,n40,n41,n42,n43,n44,n45}. 

It is the goal of the present paper to investigate the perturbations and the QNMs of the exact Schwarzschild-de Sitter type solution of the $f(R)$ gravity theory, as obtained and initially investigated in  \cite{s0} and \cite{s0a}, respectively. In our analysis we consider the scalar, vector and {axial} tensor perturbations of the $f(R)$ black hole geometry.  To obtain the quasinormal modes (QNMs) for all three types of perturbations we  use the higher-order WKB approximation and the analytical method as formulated in \cite{mashhoon_2,mashhoon_1}. For appropriately chosen parameter values, the QNMs of the $f(R)$ gravity black hole have values close to those of the standard Schwarzschild black hole. Significant differences in the physical properties of the two solutions still appear,  since the model parameter $A$ leads to a nonlinear increase in both the oscillation frequency and the damping rate of ringdown gravitational waves. A similar effect can be observed for the parameter $\eta$ for small values of $A$. On the other hand, for larger values of $A$ we find an opposite effect, with the scalar perturbations presenting a distinct behavior as compared to the vector and {axial} tensor cases. Contour plots of QNMs in the parameter space are also presented. Time-domain profiles are also obtained for the presentation of the dynamical evolution of the gravitational wave signal. 

The present paper is organized as follows. We present the generalized Schwarzschild-de Sitter spherically symmetric solution of the vacuum field equations of the $f(R)$ gravity theory in Section~\ref{sec2}. The scalar, vector and {axial} tensor perturbations of the $f(R)$ black hole geometry are obtained in Section~\ref{sec3}. The quasinormal modes of the black hole are investigated in Section~\ref{sec4} by using both the higher-order Wentzel-Kramers-Brillouin (WKB) approximation, and the analytical method proposed in \cite{mashhoon_2,mashhoon_1}.  The time domain profiles of the three perturbations are analyzed in Section~\ref{sec5}. Finally, we discuss and conclude our results in Section~\ref{sec6}.

Throughout the whole paper, we  use the natural system of units with $G=c=1$.

\section{Schwarzschild-de Sitter type solution in $f(R)$ gravity}\label{sec2}

In the $f(R)$ gravity models the gravitational action is given by \cite{Bu,Ke,Ke1,Ba}
\begin{equation}
S=\int f(R)\sqrt{-g}d^{4}x,
\end{equation}
where $f(R)$ is an arbitrary functions of the Ricci scalar $R$. Since we are
only interested in the vacuum case we do not add a matter Lagrangian to the
action.

Varying the action with respect to the metric $g_{\mu \nu }$ yields the
field equations, given by
\begin{equation}
F(R)R_{\mu \nu }-\frac{1}{2}f(R)g_{\mu \nu }-\nabla _{\mu }\nabla _{\nu
}F(R)+g_{\mu \nu }\square F(R)=0,  \label{field}
\end{equation}
where we have denoted $F(R)=df(R)/dR$. Note that the covariant derivative of
these field equations vanishes for all $f(R)$ by means of the generalized
Bianchi identities \cite{Koivisto, Bertolami}.

\subsection{Vacuum field equations in $f(R)$ gravity}

In the following we will restrict our study of $f(R)$ gravity models to the
static and spherically symmetric metric given by
\begin{equation}
ds^{2}=-e^{\nu (r)}dt^{2}+e^{\lambda (r)}dr^{2}+r^{2}\left( d\theta
^{2}+\sin ^{2}\theta d\phi ^{2}\right) .  \label{metr1}
\end{equation}
For a metric of the form given by Eq.~(\ref{metr1}), the vacuum
field equations of $f(R)$ gravity can be expressed as \cite{s0,s0a}
\begin{equation}\label{f1}
\frac{F^{\prime \prime }}{F}-\frac{1}{2}(\nu ^{\prime }+\lambda ^{\prime })%
\frac{F^{\prime }}{F}-\frac{\nu ^{\prime }+\lambda ^{\prime
}}{r}=0,
\end{equation}
\bea\label{f2}
\nu ^{\prime \prime }&+&\nu ^{\prime }{}^{2}-\frac{1}{2}(\nu ^{\prime
}+\lambda ^{\prime })\left(\nu ^{\prime }+\frac{2}{r}\right)-\frac{2}{r^{2}}(1-e^{\lambda })\nonumber\\
&&=-2
\frac{F^{\prime \prime }}{F}+\left(\lambda ^{\prime
}+\frac{2}{r}\right)\frac{F^{\prime }}{F},
\eea
\begin{equation}\label{f3}
\frac{f}{F}e^{\lambda }=-\nu ^{\prime \prime }-\frac{1}{2}(\nu
^{\prime }-\lambda ^{\prime })\nu ^{\prime }+\frac{2}{r}\lambda
^{\prime }+\left(\nu ^{\prime }+\frac{4}{r}\right)\frac{F^{\prime
}}{F},
\end{equation}
\begin{equation}\label{f4}
R=2\frac{f}{F}-3e^{-\lambda }\left[ \frac{F^{\prime \prime }}{F}+\frac{1}{2}%
\left(\nu ^{\prime }-\lambda ^{\prime
}+\frac{4}{r}\right)\frac{F^{\prime }}{F}\right] ,
\end{equation}
where the prime denotes differentiation with respect to $r$, and we have denoted $F(R)=df(R)/dR$
We introduce now a new independent variable $\sigma $ by means of the transformation
$\sigma =\ln r $. Therefore the field equations Eqs.~(\ref{f1}) and~(\ref{f2})
become \cite{s0a}
\begin{equation}
\frac{F_{,\left.\right.\sigma \sigma }}{F}-\left[ 1+\frac{1}{2}(\nu
_{,\left.\right.\sigma }+\lambda _{,\left.\right.\sigma })\right]
\frac{F_{,\left.\right.\sigma }}{F}-(\nu _{,\left.\right.\sigma
}+\lambda _{,\left.\right.\sigma })F=0\,,
\end{equation}
\bea
\nu _{,\left.\right.\sigma \sigma }&-&\nu _{,\left.\right.\sigma }+\nu
_{,\left.\right.\sigma }^{2}-\frac{1}{2}(\nu _{,\left.\right.\sigma
}+\lambda _{,\left.\right.\sigma })(\nu _{,\left.\right.
\sigma}+2)+2(1-e^{\lambda }) \nonumber\\
&=&-2\frac{F_{,\left.\right.\sigma \sigma
}}{F}+(\lambda _{,\left.\right.\sigma }+4)\frac{F_{,\left.\right.\sigma
}}{F}\,,
\eea
where the comma denotes now the differentiation with respect to the new independent variable $\sigma $.

Due to the introduction of the above coordinate transformation  the
basic field equations Eqs.~(\ref{f1}) and~(\ref{f2}) do not depend anymore
on the coordinate $\sigma $. To further simplify the mathematical formalism  we represent
the function $F$ as $F_{,\left.\right.\sigma
}/F=v$, giving $F_{,\left.\right.\sigma \sigma }/F=v_{,\left.\right.\sigma
}+v^{2}$, where $v$ is a function of the independent variable $\sigma $. Hence Eq.~(\ref{f1})
becomes
\begin{equation}\label{h1}
v_{,\left.\right.\sigma }+v^{2}-\left[ 1+\frac{1}{2}(\nu
_{,\left.\right.\sigma }+\lambda _{,\left.\right.\sigma })\right] v-(\nu
_{,\left.\right.\sigma }+\lambda _{,\left.\right.\sigma })=0.
\end{equation}

This equation is a Riccati type first order differential equation.
By using the representation based on the  function $v$ and the independent variable $\sigma$, Eq. (\ref{f2}) becomes
\bea\label{h2}
\nu _{,\left.\right.\sigma \sigma }&-&\nu _{,\left.\right.\sigma }+\nu
_{,\left.\right.\sigma }^{2}-\frac{1}{2}(\nu _{,\left.\right.\sigma
}+\lambda _{,\left.\right.\sigma })(\nu _{,\left.\right.\sigma
}+2)+2(1-e^{\lambda })\nonumber\\
&=&-2v_{,\left.\right.\sigma }-2v^{2}+(\lambda
_{,\left.\right.\sigma }+4)v.
\eea

After substituting, with the use of Eq.~(\ref{h1}), the term $v_{,\left.\right.\xi }+v^{2}$ in
Eq.~(\ref{h2}), we find
\bea\label{sol2}
2(1-e^{\lambda })&+&\nu _{,\left.\right.\sigma \sigma }-\nu
_{,\left.\right.\sigma }+\nu _{,\left.\right.\sigma }^{2}+(\nu
_{,\left.\right.\sigma }+\lambda _{,\left.\right.\sigma
})(1-\frac{1}{2}\nu _{,\left.\right.\sigma })\nonumber\\
&&-(2-\nu
_{,\left.\right.\sigma })v=0.
\eea

Once the general solution of Eqs.~(\ref{h1}) and~(\ref{sol2}) is known, we obtain the
general static vacuum solution of the field equations  of the $f(R)$ gravity theory. With $\nu (\sigma )$ and $\lambda (\sigma
)$ specified, we can obtain $v$ and then the function $F$, together with all the other relevant geometrical
quantities. If  $F$ and the metric tensor components
are known, the function $f$ can be obtained as a function of $R$ from
Eqs.~(\ref{f3}) and~(\ref{f4}) in a parametric form, as $f=f\left(
\sigma \right) $, $R=R\left( \sigma \right) $.

A general solutions of the field equations with the metric tensor components satisfying the condition  
\be\label{cond}
\nu _{,\left.\right.\sigma
}+\lambda _{,\left.\right.\sigma }=0,
\ee
can be easily obtained. We call these types of solutions  as generalized
Schwarzschild-de Sitter type solutions. By re-scaling the time coordinate the constant of integration in Eq.~(\ref{cond}) can be
taken as zero. Thus, without any loss of generality we can consider that the generalized Schwarzschild-de Sitter solution satisfies the condition $\nu
+\lambda =0 $. 

Therefore, for generalized
Schwarzschild-de Sitter geometries, Eq.~(\ref{h1}) reduces to
\be
v_{,\left.\right.\sigma }+v^{2}-v=0, 
\ee
which has the general solution
\begin{equation}
v=\frac{e^{\sigma }}{e^{\sigma }+C}=\frac{r}{r+C}\,,
\end{equation}
where $C$ is an arbitrary constant of integration. By taking into account the definition
of $v$, and by reintroducing the $r$ coordinate, we find
\be
F(r)=Ar+B, 
\ee
where $A$ and $B$ are arbitrary constants of integration. Finally, the solution of
Eq. (\ref{sol2}) gives the generalized Schwarzschild-de Sitter type vacuum
solution of $f(R)$ gravity theory, which has the form \cite{s0,s0a}
\begin{eqnarray}
e^{\nu }&=&e^{-\lambda }=1-\frac{AC_{2}}{2B^{2}}+\frac{C_{2}}{3Br}-\frac{%
A(B^{2}-AC_{2})}{B^{3}}r\nonumber\\
&&+\frac{1}{B^{2}}\left( A^{2}-
\frac{C_{1}}{6B^{2}}\right) r^{2}\nonumber\\
&&+ 
\frac{A^{2}(B^{2}-AC_{2})r^{2}}{B^{4}}\ln \left( A+\frac{B}{r}%
\right),
\end{eqnarray}
where $C_{1}$ and $C_{2}$ are arbitrary constants of integration.

There are several interesting differences between to the vacuum solutions of standard general
relativity and the present solution of the $f(R)$ gravity theory. These differences are related to the presence of two new terms in the solution, the linear term with respect to $r$,
and a term with the logarithmic dependence of $r$. Moreover, in the
general case the generalized Schwarzschild - de Sitter type exact vacuum solution in
$f(R)$ gravity depends on a set of four arbitrary integration constants
$\left(A, B, C_{1},C_{2}\right)$. 

To reobtain
the Schwarzschild-de Sitter solution of standard general
relativity we adopt the following values for the constants: $A=0$, $%
C_{2}=-6BM$ and $C_{1}=2B^{4}\Lambda $, respectively, where $M$ is
the mass of the central object \cite{s0,s0a}.

In the following  we neglect the possible effect of a cosmological
constant on the physical and geometrical properties of the black hole by taking $C_{1}=0$. After denoting  $A/B=1/\eta $,
the generalized Schwarzschild-de Sitter vacuum solution in $f(R)$ gravity is obtained in the form
\bea\label{metreta}
\hspace{-0.5cm}e^{\nu }&=&e^{-\lambda }=1+\frac{3M}{\eta }-\frac{2M}{r}-\left( 1+\frac{6M}{%
\eta }\right) \left( \frac{r}{\eta }\right)\nonumber\\
\hspace{-0.5cm}&& +\left( \frac{r}{\eta }\right)
^{2}\left\{ 1+\left( 1+\frac{6M}{\eta }\right) \ln \left[ A\left( 1+\frac{%
\eta }{r}\right) \right] \right\}. 
\eea

\subsection{Horizon and Asymptotic Structure}

To properly define the regions of interest for our quasinormal mode analysis, it is essential to characterize the asymptotic structure and the horizons of the generalized Schwarzschild-de Sitter geometry. The asymptotic behavior of the spacetime is governed by the metric function $e^{\nu(r)}$ evaluated at large radial distances.

In the limit $r \rightarrow \infty$, and assuming $r \gg \eta$ and $r \gg M$, the leading-order term of the metric coefficient $g_{tt} = e^{\nu(r)}$ from Eq. \eqref{metreta} is dominated by the quadratic radial term:
$$g_{tt} \approx \left(\frac{r}{\eta}\right)^2 \left[ 1 + \left(1 + \frac{6M}{\eta}\right) \ln A \right].$$

Let us define the parameter $\Gamma$ as the factor controlling this asymptotic behavior:
$$\Gamma \equiv 1 + \left(1 + \frac{6M}{\eta}\right) \ln A.$$

Because the parameters $\eta$ and $M$ are strictly positive, the sign of $\Gamma$ depends primarily on the dimensionless parameter $A$, and it strictly dictates the asymptotic nature of the black hole solution:

de Sitter-type (dS) regime ($\Gamma < 0$): For sufficiently small values of $A > 0$, the logarithmic term $\ln A$ is strongly negative, rendering $\Gamma < 0$. In this regime, the leading term is negative, mirroring the functional form of an asymptotically de Sitter spacetime. Consequently, the spacetime possesses a cosmological horizon $r_c$ at some large finite distance where $g_{tt}(r_c) = 0$.

Anti-de Sitter-type (AdS) regime ($\Gamma > 0$): As $A$ increases, $\Gamma$ transitions to a positive value. Here, the metric coefficient grows to $+\infty$ as $r \rightarrow \infty$, mimicking an asymptotically anti-de Sitter spacetime without a cosmological horizon.

The critical value of $A$ that separates these two geometric regimes is given by:
$$A_{crit} = \exp\left[ - \left( 1 + \frac{6M}{\eta} \right)^{-1} \right].$$

For instance, taking the parameters used in the effective potential profiles (e.g., $\eta = 50$, $M = 1$), the critical value is $A_{crit} \approx 0.409$. Thus, for $A \in \{0.001, 0.01, 0.1\}$, the spacetime is strictly dS-type, whereas for $A = 1.0$, it transitions to an AdS-type geometry. 

For the application of the standard WKB approximation and the Mashhoon method used in this study, the presence of a cosmological horizon is required so that the effective potentials vanish at both boundary turning points (the event horizon and the cosmological horizon). Therefore, throughout our numerical quasinormal mode calculations, we restrict the parameter space exclusively to the $\Gamma < 0$ (dS-type) regime. This restriction ensures the existence of a cosmological horizon and guarantees the well-posedness of the standard purely ingoing and purely outgoing boundary conditions.

\begin{figure*}[htbp!]
\begin{center}
   \includegraphics[scale = 0.62]{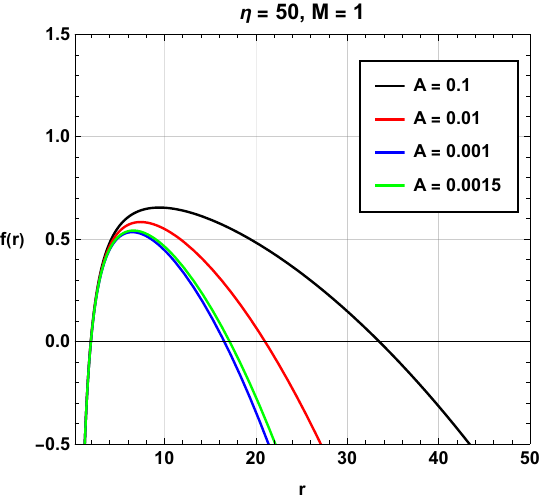}\hspace{0.5cm}
   \includegraphics[scale = 0.62]{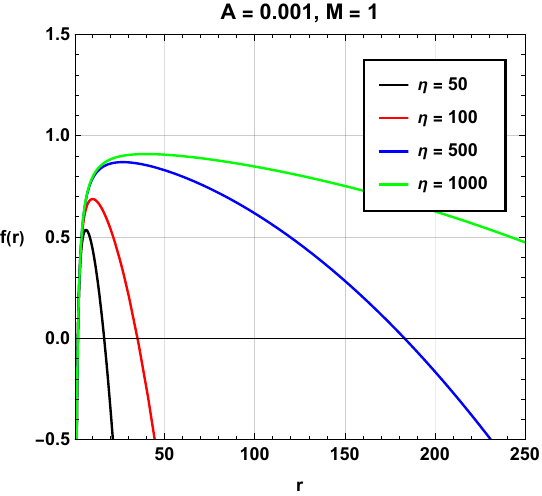}
   \end{center}
\caption{Variation of the metric function with respect to the radial coordinate  $r$. }\label{metplot}
\end{figure*}

In Fig. \ref{metplot}, we have shown the variation of the metric function for different values of the model parameters.

\section{Perturbations of the generalized Schwarzschild-de Sitter $f(R)$ black hole spacetime }\label{sec3}

In this Section, we will study three types of perturbations (scalar, electromagnetic or vector and gravitational or {axial} tensor) of the black hole geometry as given by the metric (\ref{metreta}). In the next Section we will obtain the quasinormal modes associated with these three perturbations by using the higher order WKB approximation method and the Mashhoon method, respectively.

\subsection{Massless Scalar Perturbations}

We now turn our attention to massless scalar perturbations propagating in the background of the generalized  black hole solution (\ref{metreta}) obtained in the framework of $f(R)$ gravity. The perturbations are governed by the minimally coupled massless Klein-Gordon equation
\begin{equation}
\square \Phi = \dfrac{1}{\sqrt{-g}} \partial_\mu \left( \sqrt{-g}\, g^{\mu\nu} \partial_\nu \Phi \right) = 0,
\end{equation}
which describes the dynamics of a scalar field $\Phi$, neglecting any backreaction on the spacetime geometry.

For static spherically symmetric metric the Klein-Gordon equation takes the explicit form
 
\bea
\dfrac{1}{r^2 \sqrt{g_{tt}^{-1} g_{rr}}} && \left[r^2 \sqrt{g_{tt} g_{rr}^{-1}}\, \Phi_{,r} \right]_{,r}
+ \dfrac{g_{tt}}{r^2 \sin\theta} \left( \sin\theta\, \Phi_{,\theta} \right)_{,\theta} \nonumber
\\ 
&&+ \dfrac{g_{tt}}{r^2 \sin^2\theta} \Phi_{,\phi\phi}
- \Phi_{,tt} = 0. 
\eea

The angular dependence of the scalar field suggests that the appropriate basis functions for separation of variables are the associated Legendre polynomials $P_l^m(\cos\theta)$, which satisfy
\bea
\dfrac{1}{\sin\theta} \dfrac{d}{d\theta} && \left( \sin\theta\, \dfrac{d P_l^m(\cos\theta)}{d\theta} \right) - \dfrac{m^2}{\sin^2\theta} P_l^m(\cos\theta) \nonumber \\ &&= -l(l+1) P_l^m(\cos\theta).
\eea

The $\phi$-dependence can similarly be separated via $\partial^2_\phi \Phi = -m^2 \Phi$. Consequently, the scalar field may be expanded as a multipole series:
\bea
&& \Phi(t, r, \theta, \phi) = \nonumber \\ && \dfrac{1}{r} \sum_{l,m} \sqrt{\dfrac{2l+1}{4\pi} \dfrac{(l-m)!}{(l+m)!}}\, \psi_s(t, r)\, e^{i m \phi} P_l^m(\cos\theta),
\eea
where $\psi_s(t, r)$ denotes the radial and temporal part of the scalar field perturbation and the index ``$s$'' labels the scalar nature of the mode.

Substituting this ansatz into the full Klein-Gordon equation and separating variables yields a Schrödinger-like equation for the radial component
\begin{equation}
\frac{d^2 \psi_s(r_*)}{dr_*^2} + \omega^2 \psi_s(r_*) = V_s(r)\, \psi_s(r_*),
\end{equation}
where $r_*$ is the tortoise coordinate.

The effective potential $V_s(r)$ governing the scalar perturbations is given by
\bea
V_s(r) &=& |g_{tt}(r)| \left[ \frac{l(l+1)}{r^2} + \right. \nonumber \\ && \left. \frac{1}{r \sqrt{|g_{tt}(r)|\, g_{rr}(r)}} \frac{d}{dr} \sqrt{|g_{tt}(r)|\, g_{rr}^{-1}(r)} \right].
\eea

Equivalently, using the shorthand notation $h(r) = \sqrt{|g_{tt}| / g_{rr}}$ { along with $g_{tt} = -g_{rr}^{-1}$}, the potential may be rewritten as
\begin{equation}
V_s(r) = h(r) \left[ \frac{l(l+1)}{r^2} + \frac{1}{r} \frac{d}{dr} h(r) \right].
\end{equation}

For the black hole spacetime considered in this work we obtain
\begin{widetext}
 \bea
V_s(r) &=& \frac{\left[r^3 (\eta +6 M) \ln \left(\frac{A (\eta +r)}{r}\right)+\eta  M \left(-2 \eta ^2-6 r^2+3 \eta  r\right)+\eta  r \left(\eta ^2+r^2-\eta  r\right)\right]}{\eta ^6 r^4 (\eta +r)}  \Bigg\{2 r^3 (\eta +6 M) (\eta +r) \ln \left(\frac{A (\eta +r)}{r}\right)  \nonumber\\ 
&& +\eta  \Bigg[\eta ^2 r \left(\left(l^2+l-1\right) r+2 M\right)+\eta ^3 (l (l+1) r+2 M)+2 r^3 (r-6 M)-6 \eta  M r^2\Bigg]\Bigg\}.
\eea   
\end{widetext}

This effective potential encapsulates the influence of spacetime geometry on the propagation of scalar perturbations. The multipole moment $l$ determines the angular structure of the mode, while the frequency $\omega$ is interpreted as the quasinormal mode (QNM) frequency. The wavefunction $\psi_s(r_*)$ thus describes the radial evolution of scalar QNMs \cite{chandrasekhar, n1, lopez2020}.

\begin{figure*}[htbp!]
\begin{center}
   \includegraphics[scale = 0.62]{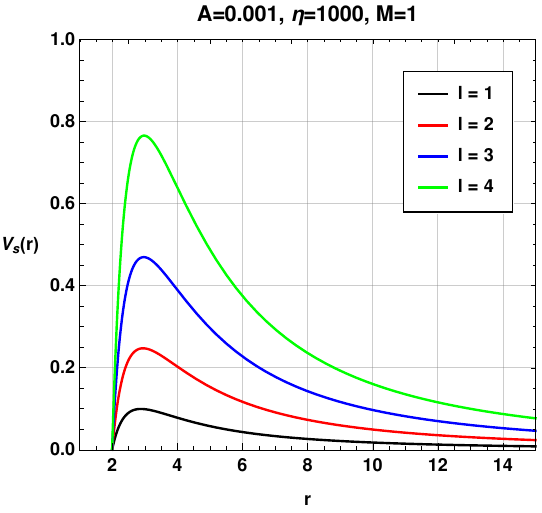}\hspace{0.5cm}
   \includegraphics[scale = 0.62]{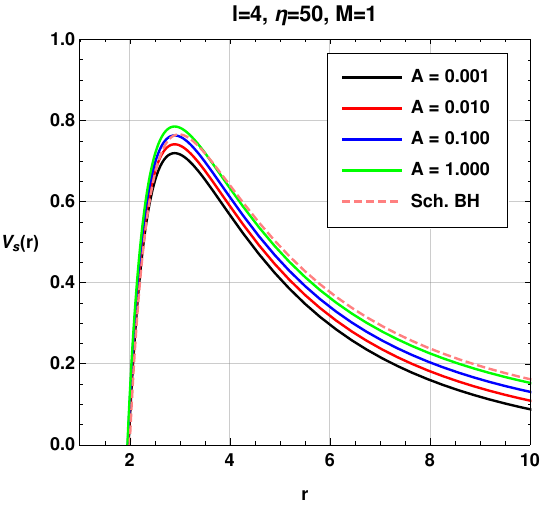}
   \includegraphics[scale = 0.62]{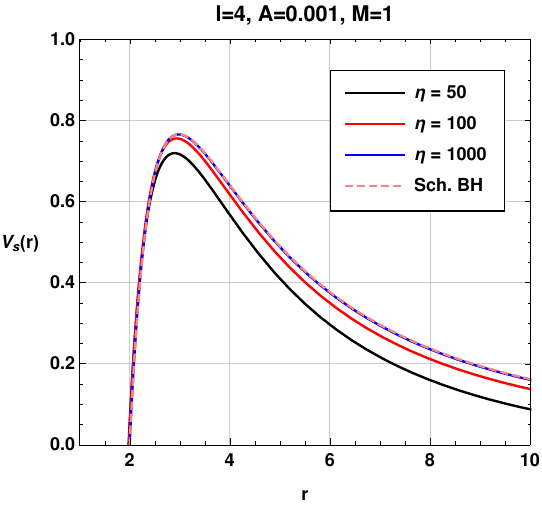} \vspace{-0.2cm}
   \end{center}
\caption{Variation of the scalar potential $V_s(r)$ as a function of the radial coordinate  $r$. }\label{fig_Vs_01}
\end{figure*}

We have shown the behaviour of the potential graphically in Fig. \ref{fig_Vs_01}.

\subsection{Vector Perturbations}
We now focus our attention on massless vector perturbations, with particular emphasis on electromagnetic field fluctuations in the context of $f(R)$ gravity. These perturbations are treated using the tetrad (vierbein) formalism, which facilitates the decomposition of the curved spacetime metric $g_{\mu\nu}$ into a locally flat Minkowski frame defined by $\eta_{\bar{\mu} \bar{\nu}}$. This formalism is especially advantageous in non-trivial geometries, as it enables the consistent projection of tensorial quantities from curved to flat spacetime via the tetrads $e_\mu^{\bar{\mu}}$, satisfying

$$
g_{\mu\nu} = e_\mu^{\bar{\mu}} \eta_{\bar{\mu} \bar{\nu}} e_\nu^{\bar{\nu}}, \quad
e^{\bar \mu}_\mu e^\mu_{\bar \nu} = \delta^{\bar \mu}_{\bar \nu}, \quad
e^{\bar \mu}_\mu e^\nu_{\bar \mu} = \delta^{\nu}_{\mu}.
$$

These relations allow one to relate curved-space vectors and tensors such as $S_\mu$ and $P_{\mu\nu}$ to their flat-space counterparts.

The dynamics of electromagnetic perturbations are encoded in the antisymmetric field strength tensor $F_{\bar{\mu} \bar{\nu}}$, which satisfies the Bianchi identity

$$
\partial_{[\bar{\mu}} F_{\bar{\nu} \bar{\rho}]} = 0,
$$

leading to a constraint structure that governs the evolution of the perturbations.

By exploiting the symmetries of a spherically symmetric background metric (typically of the form $ds^2 = -g_{tt}(r) dt^2 + g_{rr}(r) dr^2 + r^2 d\Omega^2$), the perturbation equations can be reduced. In particular, the dynamical equation for the perturbed field component $\mathcal{F} = F_{\bar{t} \bar{\phi}} \sin\theta$ takes the form
\bea
    \left[ \sqrt{g_{tt} g_{rr}^{-1}} \left( r \sqrt{g_{tt}}\, \mathcal{F} \right)_{,r} \right]_{,r} &+& \dfrac{g_{tt} \sqrt{g_{rr}}}{r} \left( \dfrac{\mathcal{F}_{,\theta}}{\sin\theta} \right)_{,\theta} \sin\theta \nonumber\\
   && - r \sqrt{g_{rr}}\, \mathcal{F}_{,tt} = 0.
\eea

In contrast to scalar perturbations, the angular part of the vector perturbation equation reveals that the relevant angular basis is furnished not by the standard spherical harmonics, but by Gegenbauer polynomials {  $C_{l+1}^{(-1/2)}(\cos\theta)$ }, which satisfy
\begin{equation}
\sin\theta \dfrac{d}{d\theta}\left(\dfrac{1}{\sin\theta} \frac{d \,{ C_{l+1}^{(-1/2)}(\cos\theta) }}{d\theta}\right)
= - l(l+1) { C_{l+1}^{(-1/2)}(\cos\theta) }.
\end{equation}
Accordingly, the total wavefunction is expanded as:

\begin{equation}
\mathcal{F}(t,r,\theta, \phi) = \sum_l \psi_{em}(t,r) \; C_{l+1}^{(-1/2)}(\cos\theta),
\end{equation}

where $\psi_{em}(t, r)$ is the radial function associated with electromagnetic perturbations.

Substituting this ansatz into the master equation and performing separation of variables yields the radial perturbation equation
\bea
    \left[ \sqrt{g_{tt} g_{rr}^{-1}} \left( r \sqrt{g_{tt}}\, \psi_{em} \right)_{,r} \right]_{,r} &+& \omega^2 r \sqrt{g_{rr}}\, \psi_{em}\nonumber\\ 
   && - \dfrac{1}{r} g_{tt} \sqrt{g_{rr}}\, l(l+1)\, \psi_{em} = 0.\nonumber\\
\eea

This form closely resembles the scalar perturbation equation but is characterized by the absence of any first-order derivative terms in the potential. The wave equation can be cast into the Schrödinger-like form using the tortoise coordinate $r_*$, defined by

$$
\dfrac{dr_*}{dr} = \sqrt{\dfrac{g_{rr}}{g_{tt}}}.
$$

With this coordinate transformation { and using $\Psi_{em}(r) = r \sqrt{g_{tt}}\, \psi_{em}(r)$, the equation reduces to:
\begin{equation}
\frac{d^2\Psi_{em}(r_*)}{dr_*^2} + \omega^2 \Psi_{em}(r_*) = V_{e}(r) \Psi_{em}(r_*),
\end{equation}
}
where the effective potential for electromagnetic perturbations is
\begin{equation}
V_{e}(r) = g_{tt}(r) \dfrac{l(l+1)}{r^2}.
\end{equation}
For our case, the potential takes the following form
\bea\label{potential}
V_{e}(r)&=&\frac{l(l+1)}{\eta ^{3}r^{3}}\Bigg[ (r^{3}(\eta +6M)\ln \left( 
\frac{A(\eta +r)}{r}\right) \nonumber\\
&&+\eta M\left( -2\eta ^{2}-6r^{2}+3\eta r\right)
+\eta r\left( \eta ^{2}+r^{2}-\eta r\right) \Bigg] . \nonumber\\
\eea

This potential profile exhibits a pure centrifugal barrier structure modified by the background geometry $h(r)$, but notably lacks the additional derivative-dependent terms present in scalar cases. The qualitative features of the potential—such as the height and width of the barrier—are directly influenced by the form of $h(r)$, which encapsulates the modifications introduced by the underlying $f(R)$ gravitational theory.

From a physical standpoint, this analysis is not only applicable to standard electromagnetic perturbations but also offers insights into the propagation of massless vector modes that arise in extended gravitational theories. In particular, within the framework of symmergent gravity and related extensions of $f(R)$ gravity, one anticipates the emergence of new massless vector fields, such as dark photons, that couple to gravity but not to the Standard Model gauge sector. These additional degrees of freedom may contribute nontrivially to gravitational wave signals, particularly through phase shifts or echoes in the ringdown phase of black hole mergers.

Furthermore, any deviations in the QNM spectrum-especially those induced by the vector sector-could serve as observational probes for testing the validity of $f(R)$ gravity or its ultraviolet completions. Thus, the study of massless vector perturbations opens a pathway to explore not only classical instabilities and stability bounds of compact objects but also quantum-gravity motivated corrections to General Relativity.

\begin{figure*}[htbp!]
\begin{center}
   \includegraphics[scale = 0.62]{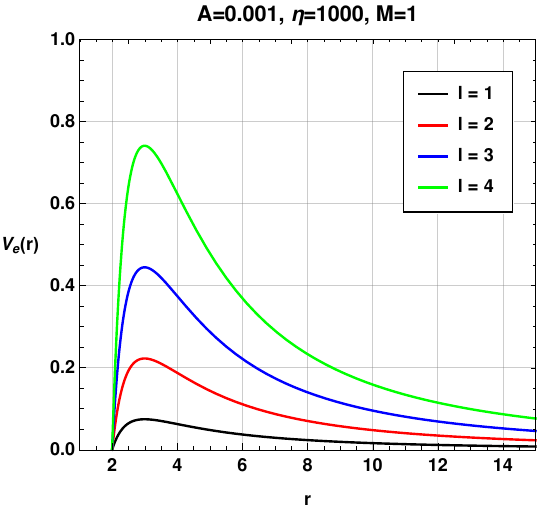}\hspace{0.5cm}
   \includegraphics[scale = 0.62]{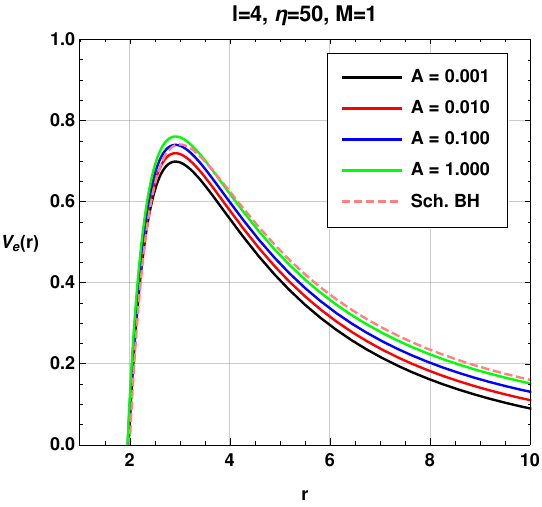}
   \includegraphics[scale = 0.62]{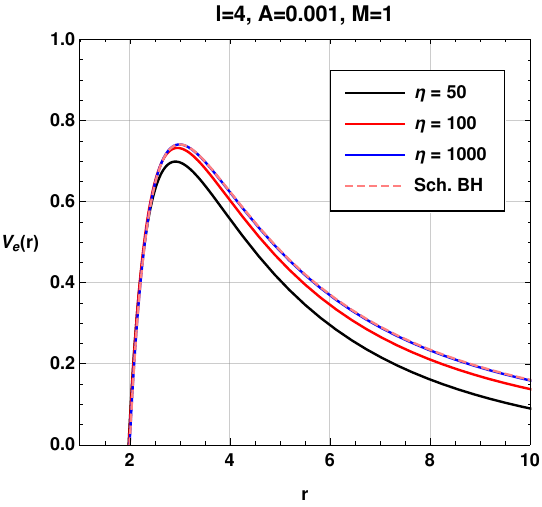} \vspace{-0.2cm}
   \end{center}
\caption{Variation of electromagnetic potential $V_e(r)$ with respect to the radial coordinate $r$. }\label{fig_Ve_01}
\end{figure*}

 Fig. \ref{fig_Ve_01}  illustrates the variation of the electromagnetic perturbation potential $V_{e}(r)$ with respect to the radial coordinate $r$, for different values of the multipole number $l$, the model parameter $A$, and the deviation parameter $\eta$, within the framework of $f(R)$ gravity.

In the top-left panel, the dependence of $V_{e}(r)$ on the angular momentum number $l$ is explored for fixed values of $A = 0.001$, $\eta = 1000$, and $M = 1$. As expected, the peak height of the potential increases monotonically with $l$. This behavior is consistent with the centrifugal barrier structure of the potential, and suggests that higher multipole modes are associated with higher potential barriers, leading to shorter-lived QNMs and higher oscillation frequencies.

The top-right panel depicts the influence of the model parameter $A$ for fixed $l = 4$, $\eta = 1000$, and $M = 1$. It is observed that increasing $A$ results in a gradual enhancement of the potential peak. Since the QNM frequencies are sensitive to the height and shape of the effective potential, this suggests that larger values of $A$ would lead to higher real parts of the QNM frequencies, possibly indicating faster oscillations, while the imaginary part (related to damping) could also be affected depending on the potential's width.

In the bottom panel, the effect of varying the parameter $\eta$ is shown for $l = 4$, $A = 0.001$, and $M = 1$. Here, an increase in $\eta$ leads to a noticeable rise in the peak of the potential. Physically, this implies that the modifications encoded in $\eta$, which may correspond to deviation scales or coupling strengths in $f(R)$ gravity, enhance the barrier potential, which again could result in QNMs with higher frequencies and reduced damping rates.

Overall, the figure highlights how the parameters $l$, $A$, and $\eta$ modulate the effective potential $V_{e}(r)$, which in turn governs the spectrum of quasi-normal modes. These dependencies are crucial in interpreting gravitational wave signals and may provide observational signatures of modifications to General Relativity inherent in the $f(R)$ framework.

\subsection{Axial Gravitational Perturbations}
\label{gravitational}

We now analyze axial (odd-parity) gravitational perturbations in the context of a static, spherically symmetric background within $f(R)$ gravity. These perturbations, characterized by their transformation under parity inversion, play a crucial role in probing the stability and dynamical behavior of compact astrophysical objects. To ensure the completeness of our study, we include {axial} gravitational perturbations alongside scalar and electromagnetic cases to compute the corresponding quasinormal modes (QNMs).

As demonstrated in Ref.~\cite{Chen:2019iuo}, for axial perturbations of an anisotropic fluid spacetime, the axial components of the perturbed energy-momentum tensor vanish. This greatly simplifies the analysis and permits the use of the tetrad formalism, reducing the Einstein field equations under perturbations to:
\begin{equation}
R_{(a)(b)} = 0,
\end{equation}
where $R_{(a)(b)}$ are the Ricci tensor components projected onto the locally flat tetrad frame.

Focusing on the $\theta\phi$ and $r\phi$ components of the perturbed field equations yields a pair of coupled equations~\cite{lopez2020}:
\be
    \left[ r^2 \sqrt{|g_{tt}|\, g_{rr}^{-1}}\,(a_{2,\theta}-a_{3,r}) \right]_{,r} = r^2 \sqrt{|g_{tt}|^{-1}\, g_{rr}}\,(a_{1,\theta}-a_{3,t})_{,t}, \label{g1}
    \ee
    \be
    \left[ r^2 \sqrt{|g_{tt}|\, g_{rr}^{-1}}\,(a_{3,r}-a_{2,\theta})\sin^3\theta \right]_{,\theta} = \frac{r^4 \sin^3\theta}{\sqrt{|g_{tt}|\, g_{rr}}}(a_{1,r}-a_{2,t})_{,t}, \label{g2}
\ee

To facilitate the separation of variables, we assume a time-harmonic dependence and decompose the perturbation function as
\begin{equation}
\mathcal{F}_g(t, r,\theta) = e^{-i\omega t} R(r) Y(\theta),
\end{equation}
where $Y(\theta)$ satisfies the angular differential equation
\begin{equation}
\frac{d}{d\theta} \left( \sin^{-3}\theta \frac{dY}{d\theta} \right) = -\left[l(l+1)-2\right]\frac{Y}{\sin^3\theta}.
\end{equation}
This angular structure is distinct from the standard spherical harmonics, with the appearance of the term $l(l+1) - 2$ reflecting the axial nature of the perturbation.

Substituting this ansatz into Eqs.~\eqref{g1} and~\eqref{g2}, and following suitable algebraic manipulations, the system reduces to a radial wave equation of Schrödinger-like form:
\begin{equation}
\frac{d^2\psi_g(r_*)}{dr_*^2} + \omega^2 \psi_g(r_*) = V_g(r)\, \psi_g(r_*),
\end{equation}
where the radial wavefunction is defined as $\psi_g(r) = R(r)/r$ and $r_*$ is the tortoise coordinate, defined through
\begin{equation}
\frac{dr_*}{dr} = \sqrt{\frac{g_{rr}}{|g_{tt}|}}.
\end{equation}

The effective potential governing axial gravitational perturbations is given by
\bea
\hspace{-0.7cm}V_g(r) &=& |g_{tt}(r)|\Bigg[\frac{2}{r^2} \left( \frac{1}{g_{rr}(r)} - 1 \right) + \frac{l(l+1)}{r^2} \nonumber\\
\hspace{-0.7cm}&&- \frac{1}{r\sqrt{|g_{tt}(r)|\, g_{rr}(r)}} \frac{d}{dr} \left( \sqrt{|g_{tt}(r)|\, g_{rr}^{-1}(r)} \right) \Bigg].
\eea
For the black hole spacetime considered here, the potential evaluates to:

    \bea
V_g(r) &=& \frac{\left[\left(l^2+l-1\right) r^2+\eta  l (l+1) r-6 \eta  M\right]}{\eta ^3 r^4 (\eta +r)} \nonumber\\
&& \times \Bigg\{r^3 (\eta +6 M) \ln \left[\frac{A (\eta +r)}{r}\right]\nonumber\\
&&+\eta  M \left(-2 \eta ^2-6 r^2+3 \eta  r\right) +\eta  r \left(\eta ^2+r^2-\eta  r\right)\Bigg\}.\nonumber\\
    \eea

This potential encapsulates the influence of spacetime curvature on axial gravitational waves and is pivotal in determining the QNM spectrum of the black hole. Its structure directly reflects the behavior of the background geometry, and deviations from the Schwarzschild or Reissner–Nordström form can significantly alter the nature of gravitational ringdown signals. We will analyze this potential numerically in the subsequent sections to extract the QNM frequencies and assess the stability properties of the solution.

\begin{figure*}[htbp!]
\begin{center}
   \includegraphics[scale = 0.62]{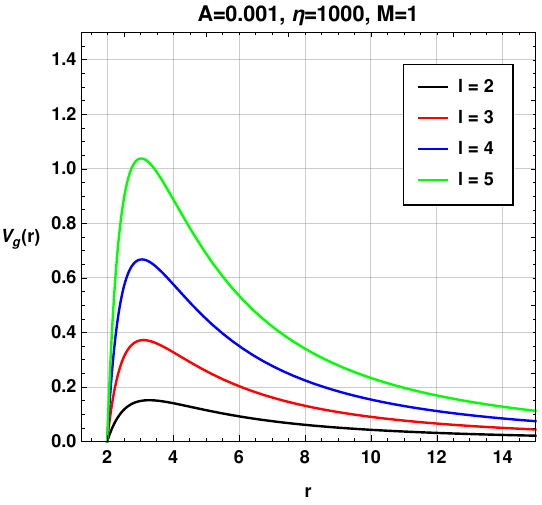}\hspace{0.5cm}
   \includegraphics[scale = 0.62]{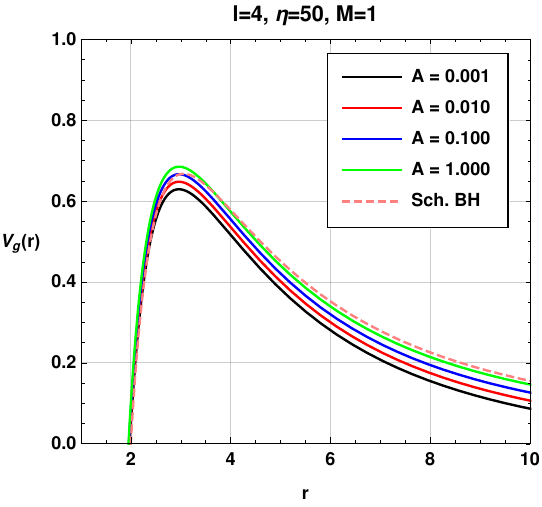}
   \includegraphics[scale = 0.62]{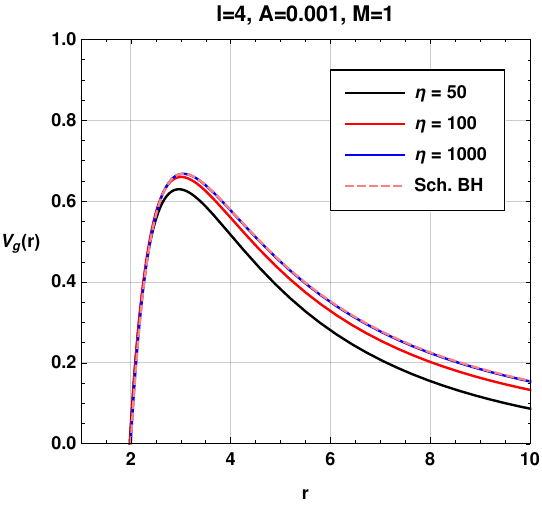} \vspace{-0.2cm}
   \end{center}
\caption{Variation of gravitational potential $V_g(r)$ w.r.t.\ $r$. }\label{fig_Vg_01}
\end{figure*}

The behaviour of the gravitational potential is shown in Fig. \ref{fig_Vg_01}.

\section{Quasinormal Modes}\label{sec4}

In this work, our objective is to determine the QNMs of black holes by applying the Wentzel-Kramers-Brillouin (WKB) approximation, a well-established method in black hole perturbation studies and the Mashhoon method. 

\subsection{The WKB approach}

First proposed by Schutz and Will \cite{Schutz}, the WKB approach offers a first-order approximation for QNM calculations. Despite its effectiveness, this method has known drawbacks, including increased errors in certain scenarios. 

To mitigate these issues, advanced higher-order WKB techniques have been developed, significantly enhancing the precision of QNM computations \cite{Will_wkb, n2, n21}. Our study leverages the improvements outlined in \cite{n21}, which integrates Padé approximations into the WKB framework to achieve superior accuracy.

Utilizing these advancements, we employ the Padé-averaged 6th-order WKB method to compute the QNMs of black holes. Previous research, such as \cite{n2}, has demonstrated that Padé averaging improves the accuracy of QNM estimates. By adopting this refined technique, we aim to obtain more precise and dependable QNM values, which we will compare with the Mashhoon method to confirm their validity. This comparison is essential for assessing the reliability of the WKB approach and deepening our insights into black hole dynamics.

The computation of QNMs for black holes is a vital tool in understanding the stability and dynamical properties of compact objects in both general relativity and its extensions, including $f(R)$ gravity. In this work, we consider three types of perturbations {\it viz.}, scalar perturbations, electromagnetic (massless vector) perturbations and {axial} gravitational perturbations of a spherically symmetric black hole in $f(R)$ gravity background and aim to obtain the QNM frequencies using the WKB approximation method and the Mashhoon method. Before applying these semi-analytical methods, it is essential to assess whether the effective potential satisfies the necessary structural and boundary conditions that underpin the validity of the WKB formalism.

One may note that for the selected ranges of the model parameters, the potentials associated with all three types of perturbations behave in a favourable way as described in the Figs. \ref{fig_Vs_01}, \ref{fig_Ve_01} and \ref{fig_Vg_01}. The potentials have well-defined single peaks and are asymptotically finite in nature.

To investigate the behaviour of the potential further, we pick the electromagnetic potential as an example.
The potential associated with electromagnetic perturbations in this background, as derived in Eq. \eqref{potential}, is a function of the radial coordinate $r$, the multipole index $l$, and model parameters such as $A$ and $\eta$. The effective potential includes a logarithmic structure and is modulated by the combination $(\eta + 6 M)\ln\left[\frac{A(\eta + r)}{r}\right]$, which contributes to a non-trivial asymptotic behaviour. 

Importantly, however, numerical and analytical inspection reveals that the potential forms a smooth single-peaked barrier for all reasonable values of the parameters. This structure is crucial for the WKB approximation, which relies on the existence of two classical turning points between which the perturbations are trapped.

To ensure the validity of the method, we analyse the asymptotic form of the potential as $r \to \infty$. Unlike standard black hole potentials in general relativity that decay to zero at spatial infinity, the potential considered here asymptotes to a constant value

\begin{equation}
 \lim_{r\rightarrow \infty}V_e(r)=V_{\infty}=   \frac{l (l+1) [(\eta +6) \ln (A)+\eta ]}{\eta ^3}.
\end{equation}

For the parameter values used in our calculations ( {\it e.g.,}$A = 0.0015$, $\eta = 1000$, $M = 1$), this evaluates to a small negative constant (e.g., $V_\infty \approx -10^{-5}$ for $l = 1$). Although this deviates from the vanishing asymptotic behaviour assumed in standard QNM analyses, the impact is minimal: the wave equation remains of Schrödinger type, and the solution at infinity is still oscillatory since $\omega^2 + |V_\infty| > 0$. 

Thus, the standard boundary conditions—purely ingoing waves at the event horizon and purely outgoing waves at spatial infinity—remain valid, justifying the continued use of the WKB approximation.

The potential barrier's peak structure also plays a critical role. As illustrated in Fig. \ref{fig_Ve_01}, the peak height increases with the multipole index $l$, consistent with the expected centrifugal barrier effect. Furthermore, variations in parameters $A$ and $\eta$ shift the peak location and height but do not alter the qualitative form of the potential. The potential remains differentiable and single-peaked, with two turning points on either side of the peak for all cases considered. These features guarantee the existence of a quasi-bound state structure, which is fundamental for the applicability of the WKB method.\\

 Before applying the WKB approximation, it is crucial to carefully address the asymptotic behavior of the effective potentials and the corresponding boundary conditions. As discussed in the previous part, the asymptotic structure of the spacetime transitions from a de Sitter-type (dS) to an anti-de Sitter-type (AdS) geometry depending on the parameter $A$. While the electromagnetic potential $V_e(r)$ remains well-behaved and asymptotes to a constant, the scalar and gravitational potentials, $V_s(r)$ and $V_g(r)$, diverge as $r \rightarrow \infty$ in the strictly AdS-like regime. In such a regime, the standard boundary conditions for quasinormal modes fail, and one must instead impose Dirichlet or Robin boundary conditions at spatial infinity, which generally renders the standard WKB method inapplicable.

However, for the dS-like regime (where the asymptotic factor $\Gamma < 0$, corresponding to small values of $A$ such as $0.001$ and $0.0015$ chosen for our numerical calculations), the spacetime possesses a well-defined cosmological horizon, $r_c$. As $r \rightarrow r_c$, the tortoise coordinate $r_* \rightarrow +\infty$ and the metric function $g_{tt} \rightarrow 0$. Consequently, all effective potentials ($V_s$, $V_e$, and $V_g$) vanish at the cosmological horizon. This ensures that the boundary conditions naturally reduce to purely ingoing waves at the black hole event horizon ($r_* \rightarrow -\infty$) and purely outgoing waves at the cosmological horizon ($r_* \rightarrow +\infty$). By strictly restricting our parameter space to this dS-like regime throughout our numerical analysis, we guarantee that the potentials smoothly approach zero (or a negligibly small constant) at the turning point boundaries, thereby fully satisfying the requirements for, and justifying the use of, the Padé-averaged 6th-order WKB approximation. To clarify this, let us pick an example using $\eta = 1000$ and $A=0.001$ in the units of M. The event horizon is located at $1.99802$ while the cosmological horizon is located at $368.123$. One can see that the scalar potential $V_s(r) =0 $ at the cosmological horizon, justifying the validity of the WKB method.
{
Furthermore, while restricting our analysis to the dS-like regime ensures the appropriate boundary conditions---purely ingoing waves at the event horizon and purely outgoing waves at the cosmological horizon---it is equally critical to confirm the structural integrity of the potentials between these boundaries. Although the metric function $h(r) \rightarrow 0$ at the cosmological horizon, one must ensure that the effective potentials do not develop pathological features. 

Numerical inspection of the scalar potential $V_s(r)$ and the axial gravitational potential $V_g(r)$ across our adopted parameter space ($\Gamma < 0$) confirms that both potentials consistently maintain a smooth, single-peaked barrier structure. There are no extra local extrema or potential wells formed outside the main peak. This smooth, single-barrier structure guarantees the existence of exactly two well-defined classical turning points, which is a fundamental prerequisite for the WKB method. Because these structural and asymptotic conditions are fully met, the Pad\'e-averaged sixth-order WKB approximation remains strictly convergent and mathematically well-posed for the scalar and axial gravitational perturbations, just as it is for the electromagnetic case.
}

 Within the framework of the sixth-order WKB approximation, the quasinormal frequencies are determined by the relation \cite{n2}
\begin{equation}
i\frac{\left(\omega^{2}-V_{\text{max}}\right)}{\sqrt{-2V_{\text{max}}^{\prime\prime}}} - \sum_{j=2}^{6}\Lambda_{j} = n + \tfrac{1}{2},
\label{fr2}
\end{equation}
where $V_{\text{max}}$ denotes the maximum of the effective potential $V(r)$, and $V_{\text{max}}^{\prime\prime} = \left.\frac{d^{2}V(r_{\ast})}{dr_{\ast}^{2}}\right|_{r_{\ast}=r_{\text{max}}}$ corresponds to the second derivative of the potential with respect to the tortoise coordinate, evaluated at the peak position $r_{\text{max}}$. The quantities $\Lambda_{j}$ represent the higher-order WKB correction terms, the explicit expressions of which can be found in Refs. \cite{Iyer,n2}. 

This formulation provides a reliable estimation of the quasinormal spectrum, particularly in the regime of low overtone numbers $n$, with the accuracy improving when the condition $n < l$ is satisfied. In comparison to lower-order WKB schemes, the sixth-order expansion significantly enhances precision, making it especially suitable for black hole potentials where higher accuracy in capturing the quasinormal modes is required.\\

We present the QNM values calculated using the Padé-averaged 6th-order WKB method, accompanied by error estimates in Tables \ref{tab01scalar}, \ref{tab01} and \ref{tab01Grav} for scalar, electromagnetic and {axial} gravitational perturbations, respectively. The third and fourth columns of the table display the errors associated with the WKB approximation, specifically the root mean square error, denoted as $\vartriangle_{rms}$, and the error term $\Delta_6$, defined as \cite{n2}
\begin{equation}\Delta_6 = \dfrac{\vline \; \omega_7 - \omega_5 \; \vline}{2},\end{equation}
where $\omega_7$ and $\omega_5$ correspond to QNMs computed using the 7th and 5th-order Padé-averaged WKB methods, respectively. 

Through this error analysis, we provide a comprehensive evaluation of the accuracy of our QNM estimates, further validating the effectiveness of the WKB method in studying black hole perturbations. The computed frequencies are presented for various values of $l$, with the overtone number fixed at $n = 0$. The results in Tables \ref{tab01scalar}, \ref{tab01} and \ref{tab01Grav} show that the real part of the frequency increases with $l$, indicating more rapid oscillations for higher multipole modes. 

Meanwhile, the imaginary part remains consistently negative, indicating the presence of damped and stable modes. Most importantly, the convergence indicators $\vartriangle_{\text{rms}}$ and $\Delta_6$ decrease rapidly with increasing $l$, confirming the accuracy and internal consistency of the WKB expansion in this regime. The tables include the QNMs for the Schwarzschild black hole, along with the percentage deviation of the obtained QNMs from the Schwarzschild case. Notably, the gravitational QNMs are closer to the corresponding QNMs of the Schwarzschild black hole.

An important consistency condition, $\omega^2 > V_\infty$, is also satisfied across all computed modes. For example, at $l = 1$, we have $\omega = 0.248247 - 0.0924829i$, leading to $|\omega|^2 \approx 0.0703$, whereas $V_\infty \approx -1.11 \times 10^{-5}$. This inequality becomes even more strongly satisfied for higher values of $l$, ensuring that the wavefunction remains oscillatory in the asymptotic regime and justifying the physical boundary conditions required for QNM analysis.

In summary, the WKB method is well-justified for analysing these perturbations in the modified black hole background considered in this study. The potential satisfies the necessary smoothness and barrier-like structure conditions, and its asymptotic behaviour, while non-zero, does not invalidate the standard QNM boundary conditions. The sixth-order WKB results show excellent convergence when combined with Padé averaging, and the consistency condition $\omega^2 > V_\infty$ is thoroughly met across all modes. Together, these observations affirm that the WKB approximation remains a powerful and valid tool for extracting QNM spectra in extended gravity theories.

\subsection{The analytical approach} 

In addition to the WKB approach for calculating QNMs, we employ an analytical technique developed in \cite{mashhoon_0, mashhoon_2,mashhoon_1}, which provides an approximate but elegant method for estimating QNMs in black hole spacetimes. This approach is based on matching the effective potential arising from perturbations to a known solvable potential, such as the Pöschl–Teller or Eckart potential. In this work, we focus on scalar field perturbations around the black hole background under consideration and apply the analytical method \cite{mashhoon_2,mashhoon_1} by comparing the resulting effective potential to the Pöschl–Teller form.

To initiate this analysis, we consider the effective potential associated with the vector perturbation.
Once the effective potential is derived, the analytical method \cite{mashhoon_2,mashhoon_1} proceeds by approximating this potential near its peak using the Pöschl–Teller potential, a well-known solvable model
\begin{equation}
V_{\text{PT}}(x) = \frac{V_0}{\cosh^2\left[a(x - x_0)\right]},
\label{eq13}
\end{equation}
where $V_0$ represents the maximum height of the potential and $a$ encodes its curvature at the peak location $x_0$. The rationale behind this approximation lies in the observation that, near the maximum, many physically motivated black hole potentials resemble the Pöschl–Teller form, particularly in the vicinity of the trapping region.

To match the two potentials, we identify the peak of $V_e(r)$ and evaluate both the value and the second derivative at that point to determine the effective parameters $V_0$ and $a$. These parameters are then substituted into the analytic expression for the QNMs of the Pöschl–Teller potential, as given in \cite{mashhoon_2,mashhoon_1}
\begin{equation}
\omega = \omega_R + i \omega_I = \pm \sqrt{V_0 - \frac{a^2}{4}} + i a\left(n + \frac{1}{2}\right),
\label{eq14}
\end{equation}
where $n$ is the overtone number, and $\omega_R$, $\omega_I$ denote the real and imaginary parts of the QNM frequency, respectively. This approach provides an analytic estimate for the QNM spectrum, which is especially useful for gaining qualitative insights into the system's response to scalar perturbations.

The Mashhoon method offers an efficient alternative to numerical or higher-order semi-analytic techniques, particularly when the effective potential is sufficiently smooth and bell-shaped. Although it is not expected to match the precision of higher-order WKB or time-domain integrations for general modes, it often produces remarkably accurate results for the fundamental QNMs (with $n = 0$) and large $l$. 

In the present context, the use of the analytical  method proposed in \cite{mashhoon_2,mashhoon_1} complements our WKB-based analysis and provides an independent check on the qualitative features of the quasi-normal spectrum arising from scalar field perturbations in the modified gravity background.
\begin{table*}
\centering
\begin{tabular}{cccccc}
\toprule
$l$ & {$\omega$ (Pad\'e averaged WKB)} & {$\vartriangle_{rms}$} & {$\Delta_6$} & $\omega ({\rm Schwarzschild})$ & $\%$ Deviation \\
\midrule
1 & $0.292806 - 0.0976995\,i$ & $1.37995 \times 10^{-6}$ & $2.34676 \times 10^{-5}$ & $0.29291\, -0.0977616\, i$ & $0.0391292\%$ \\
2 & $0.483559 - 0.0967725\,i$ & $5.88390 \times 10^{-7}$ & $2.41044 \times 10^{-6}$ & $ 0.483642\, -0.0967661\, i$ & $0.0168545\% $ \\
3 & $0.675299 - 0.0965057\,i$ & $9.76689 \times 10^{-8}$ & $4.63489 \times 10^{-7}$ & $ 0.675366\, -0.0965006\, i$ & $0.00985323\% $\\
4 & $0.867356 - 0.0963946\,i$ & $2.58362 \times 10^{-8}$ & $1.30077 \times 10^{-7}$ & $0.867416 - 0.0963919 \,i $ & $0.00683661\% $\\
5 & $1.059560 - 0.0963381\,i$ & $9.01367 \times 10^{-9}$ & $1.36794 \times 10^{-8}$ & $1.05961\, -0.0963368\, i $ & $0.00487106\% $\\
6 & $1.251830 - 0.0963055\,i$ & $3.73045 \times 10^{-9}$ & $5.91078 \times 10^{-9}$ & $1.25189\, -0.0963051 \,i $ & $0.00450213\% $\\
7 & $1.444150 - 0.0962851\,i$ & $1.73734 \times 10^{-9}$ & $2.84101 \times 10^{-9}$ & $ 1.44421\, -0.0962852\, i$ & $0.00401993\% $\\
8 & $1.636510 - 0.0962714\,i$ & $8.94251 \times 10^{-10}$ & $1.53125 \times 10^{-9}$ & $1.63656\, -0.0962719 \,i $ & $ 0.00305925\%$\\
\bottomrule
\end{tabular}
\caption{The scalar QNMs for $n= 0$, $M=1$, $A = 0.0015$ and $\eta=1000$ using the 6th order WKB approximation method averaged with Pad\'e approximants.}
\label{tab01scalar}
\end{table*}
\begin{table*}
\centering
\begin{tabular}{cccccc}
\toprule
$l$ & {$\omega$ (WKB)} \;\;\;\; & {$\omega$ ({\rm analytical})} \;\; & {Abs \% Diff} & {\% Dev Re} & {\% Dev Im} \\
\midrule
1 & $0.292806 - 0.097699\,i$ & $0.298704 - 0.099817\,i$ & 2.03032 & 2.01430 & 2.16752 \\
2 & $0.483559 - 0.096772\,i$ & $0.487483 - 0.096923\,i$ & 0.796416 & 0.81160 & 0.15593 \\
3 & $0.675299 - 0.096506\,i$ & $0.678175 - 0.096076\,i$ & 0.426286 & 0.42589 & 0.44524 \\
4 & $0.867356 - 0.096395\,i$ & $0.869616 - 0.095723\,i$ & 0.270094 & 0.26051 & 0.69637 \\
5 & $1.059560 - 0.096338\,i$ & $1.061414 - 0.095544\,i$ & 0.189899 & 0.17534 & 0.82409 \\
6 & $1.251830 - 0.096306\,i$ & $1.253409 - 0.095441\,i$ & 0.143203 & 0.12593 & 0.89768 \\
7 & $1.444150 - 0.096285\,i$ & $1.445523 - 0.095376\,i$ & 0.113512 & 0.09477 & 0.94390 \\
8 & $1.636510 - 0.096271\,i$ & $1.637715 - 0.095333\,i$ & 0.093362 & 0.07388 & 0.97481 \\
\bottomrule
\end{tabular}
\caption{Comparison of scalar QNMs using WKB and the analytical \cite{mashhoon_2,mashhoon_1} methods ($n= 0$, $M=1$, $A = 0.0015$ and $\eta=1000$).} \label{tab02scalar}
\end{table*}
\begin{table*}
\centering
\begin{tabular}{cccccc}
\toprule
$l$ & {$\omega$ (Pad\'e averaged WKB)} & {$\vartriangle_{rms}$} & {$\Delta_6$}& $\omega ({\rm Schwarzschild})$ & $\%$ Deviation \\
\midrule
1 & $0.248247 - 0.0924829\,i$ & $8.9582 \times 10^{-6}$ & $7.4557 \times 10^{-6}$ & $0.248191 - 0.092637 \,i$ & $0.0618467\%$ \\
2 & $0.457586 - 0.0950030\,i$ & $5.6605 \times 10^{-7}$ & $6.3003 \times 10^{-7}$ & $0.457593 - 0.0950112  \,i$ &  $0.00235685 \%$\\
3 & $0.656886 - 0.0956144\,i$ & $9.7883 \times 10^{-8}$ & $1.2131 \times 10^{-7}$ & $ 0.656898 - 0.0956171 \,i$ &  $0.00192189 \%$\\
4 & $0.853079 - 0.0958581\,i$ & $2.6896 \times 10^{-8}$ & $3.5517 \times 10^{-8}$ & $ 0.853095 - 0.0958601  \,i$ &  $ 0.00189698\%$\\
5 & $1.047890 - 0.0959798\,i$ & $9.3651 \times 10^{-9}$ & $1.3261 \times 10^{-8}$ & $ 1.04791 - 0.0959817 \,i$ &  $ 0.0021719\%$\\
6 & $1.241970 - 0.0960493\,i$ & $3.9365 \times 10^{-9}$ & $5.6829 \times 10^{-9}$ & $ 1.242 - 0.0960512  \,i$ &  $ 0.0020847\%$\\
7 & $1.435610 - 0.0960928\,i$ & $3.4818 \times 10^{-9}$ & $2.8265 \times 10^{-9}$ & $1.43564 - 0.0960947 \,i$ &  $ 0.00215435\%$\\
8 & $1.628970 - 0.0961218\,i$ & $1.6400 \times 10^{-9}$ & $1.4870 \times 10^{-9}$ & $ 1.629 - 0.0961236 \,i$ &  $ 0.00205873\%$\\
\bottomrule
\end{tabular}
\caption{The electromagnetic QNMs for $n= 0$, $M=1$, $A = 0.0015$ and $\eta=1000$ using the 6th order WKB approximation method averaged with Pad\'e approximants.}
\label{tab01}
\end{table*}
\begin{table*}
\centering
\begin{tabular}{cccccc}
\toprule
$l$ & {$\omega$ (WKB)} \;\;\;\; & {$\omega$ ({\rm analytical})} \;\; & {Abs \% Diff} & {\% Dev Re} & {\% Dev Im} \\
\midrule
1 & $0.248247 - 0.092474 i$ & $0.254974 - 0.095181 i$ & 2.73731 & 2.70995 & 2.91695 \\
2 & $0.457586 - 0.095002 i$ & $0.461686 - 0.095181 i$ & 0.878104 & 0.89598 & 0.18688 \\
3 & $0.656886 - 0.095614 i$ & $0.659824 - 0.095181 i$ & 0.447405 & 0.44727 & 0.45379 \\
4 & $0.853079 - 0.095858 i$ & $0.855367 - 0.095181 i$ & 0.277991 & 0.26823 & 0.70682 \\
5 & $1.04789 - 0.095980 i$ & $1.04977 - 0.095181 i$ & 0.193563 & 0.17878 & 0.83274 \\
6 & $1.24197 - 0.096049 i$ & $1.24356 - 0.095181 i$ & 0.145161 & 0.12769 & 0.90453 \\
7 & $1.43561 - 0.096093 i$ & $1.43699 - 0.095181 i$ & 0.114670 & 0.09576 & 0.94935 \\
8 & $1.62897 - 0.096122 i$ & $1.63019 - 0.095181 i$ & 0.094099 & 0.07448 & 0.97920 \\
\bottomrule
\end{tabular}
\caption{Comparison of electromagnetic QNMs using WKB and the analytical \cite{mashhoon_2,mashhoon_1} methods ($n= 0$, $M=1$, $A = 0.0015$ and $\eta=1000$).} \label{tab02}
\end{table*}
\begin{table*}
\centering
\begin{tabular}{cccccc}
\toprule
$l$ & {$\omega$ (Pad\'e averaged WKB)} & {$\vartriangle_{rms}$} & {$\Delta_6$} & $\omega ({\rm Schwarzschild})$ & $\%$ Deviation  \\
\midrule
2 & $0.3735978 - 0.0889510\, i$ & $2.2177 \times 10^{-6}$ & $2.3036 \times 10^{-5}$ & $0.373619 - 0.088891 \, i$ & $0.016606 \%$\\
3 & $0.5994282 - 0.0927107\, i$ & $5.9669 \times 10^{-8}$ & $5.8114 \times 10^{-7}$ & $ 0.599443 - 0.0927025 \, i$ & $ 0.00284122 \%$\\
4 & $0.8091613 - 0.0941676\, i$ & $4.6246 \times 10^{-8}$ & $1.5182 \times 10^{-8}$ & $ 0.809178 - 0.0941641 \, i$ & $ 0.00213728 \%$\\
5 & $1.0122749 - 0.0948722\, i$ & $1.8388 \times 10^{-8}$ & $9.8612 \times 10^{-9}$ & $ 1.0123 - 0.0948706 \, i$ & $ 0.00201325 \%$\\
6 & $1.2119858 - 0.0952664\, i$ & $7.2942 \times 10^{-9}$ & $4.6274 \times 10^{-9}$ & $ 1.21201 - 0.0952659 \, i$ & $ 0.00197606 \%$\\
7 & $1.4097074 - 0.0955096\, i$ & $3.2026 \times 10^{-9}$ & $2.5376 \times 10^{-9}$ & $ 1.40974 - 0.0955096  \, i$ & $ 0.0019618 \%$\\
8 & $1.6061623 - 0.0956702\, i$ & $1.5422 \times 10^{-9}$ & $1.3675 \times 10^{-9}$ & $1.60619 - 0.0956707  \, i$ & $ 0.00195346 \%$\\
\bottomrule
\end{tabular}
\caption{The gravitational QNMs for $n= 0$, $M=1$, $A = 0.0015$ and $\eta=1000$ using the 6th order WKB approximation method averaged with Pad\'e approximants.}
\label{tab01Grav}
\end{table*}
\begin{table*}
\centering
\begin{tabular}{cccccc}
\toprule
$l$ & {$\omega$ (WKB)} \;\;\;\; & {$\omega$ ({\rm analytical})} \;\; & {Abs \% Diff} & {\% Dev Re} & {\% Dev Im} \\
\midrule
2 & $0.373598 - 0.0889510 i$ & $0.378439 - 0.0897478 i$ & 1.27748 & 1.28159 & 0.88786 \\
3 & $0.599428 - 0.0927107 i$ & $0.602562 - 0.0923335 i$ & 0.52040 & 0.52084 & 0.40889 \\
4 & $0.809161 - 0.0941676 i$ & $0.811536 - 0.0934915 i$ & 0.30312 & 0.29242 & 0.72357 \\
5 & $1.01227 - 0.0948722 i$ & $1.01419 - 0.0940651 i$ & 0.20483 & 0.18909 & 0.85732 \\
6 & $1.21199 - 0.0952664 i$ & $1.21360 - 0.0943889 i$ & 0.15109 & 0.13279 & 0.93084 \\
7 & $1.40971 - 0.0955096 i$ & $1.41110 - 0.0945893 i$ & 0.11814 & 0.09870 & 0.97119 \\
8 & $1.60616 - 0.0956702 i$ & $1.60739 - 0.0947220 i$ & 0.09630 & 0.07656 & 0.99904 \\
\bottomrule
\end{tabular}
\caption{Comparison of gravitational QNMs using WKB and the analytical \cite{mashhoon_2,mashhoon_1}  methods ($n= 0$, $M=1$, $A = 0.0015$ and $\eta=1000$).} \label{tab02Grav}
\end{table*}

To examine the reliability and consistency of different semi-analytic techniques for calculating QNMs, we present a comparative study of the sixth-order WKB approximation and the analytic method \cite{mashhoon_2,mashhoon_1} in Tables \ref{tab02scalar}, \ref{tab02} and \ref{tab02Grav} for scalar, electromagnetic and {axial} gravitational perturbations. This comparison is performed for fundamental ($n = 0$) modes, using a black hole background characterised by $A = 0.0015$, $\eta = 1000$, and $M = 1$. The multipole index $l$ ranges from 1 to 8 (for {axial} gravitational perturbation, it ranges from 2 to 8), and the goal is to analyse how closely the two methods agree as $l$ increases.

A key aspect of the WKB approximation is that its accuracy depends significantly on the value of $l - n$. For cases where $l \gg n$, the method provides highly accurate results, as the potential barrier becomes steeper and better approximated by a single turning-point structure with well-defined classical turning points. Conversely, when $l$ is close to $n$, such as in the case of low $l$ values and the fundamental mode $n = 0$, the approximation deteriorates due to the shallower nature of the potential and the less pronounced barrier shape. This behaviour is clearly reflected in the deviations between the WKB and the analytic results.

Interestingly, while the analytic method \cite{mashhoon_2,mashhoon_1} slightly overestimates both the real and imaginary components of the frequency, the differences are consistently small and decrease smoothly with increasing $l$. The real part, which corresponds to the oscillation frequency, is particularly well captured by both methods from moderate $l$ onwards. The imaginary part, related to the damping rate, shows a marginally higher deviation, which can be attributed to the fact that the curvature of the effective potential at the peak—parameterized by $a$ in the Pöschl–Teller approximation—is more sensitive to small inaccuracies in the potential matching.

Overall, both methods agree to within a few percent even at low multipole values and converge rapidly as $l$ increases. The maximum deviation is under $3\%$, and for $l \geq 3$, the difference falls below $0.5\%$. This excellent agreement in the high $l$ regime not only confirms the internal consistency of the perturbative analysis but also validates the structural form of the effective potential derived from the electromagnetic perturbations in the modified gravity background. Importantly, it affirms that while the WKB method may lose precision at low $l$, it remains highly trustworthy for higher multipole indices, particularly when $l - n \gg 1$, which is typically the case in astrophysical applications.

Thus, Tables \ref{tab02scalar}, \ref{tab02} and \ref{tab02Grav} for scalar, electromagnetic and {axial} gravitational perturbations demonstrate that the WKB and Mashhoon methods yield closely matching results for QNMs, especially when the condition $l \gg n$ is satisfied. The improved accuracy of WKB in the high-$l$ regime and the analytic convenience of the Mashhoon method make them complementary tools for studying black hole spectroscopy in extended gravity theories.

\begin{figure*}[htbp!]
\begin{center}
   \includegraphics[scale = 0.50]{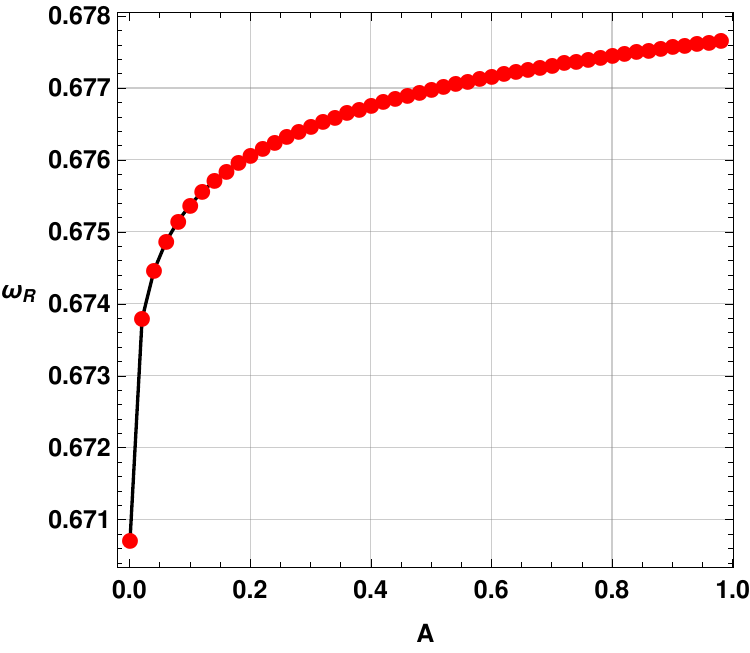}\hspace{0.5cm}
   \includegraphics[scale = 0.50]{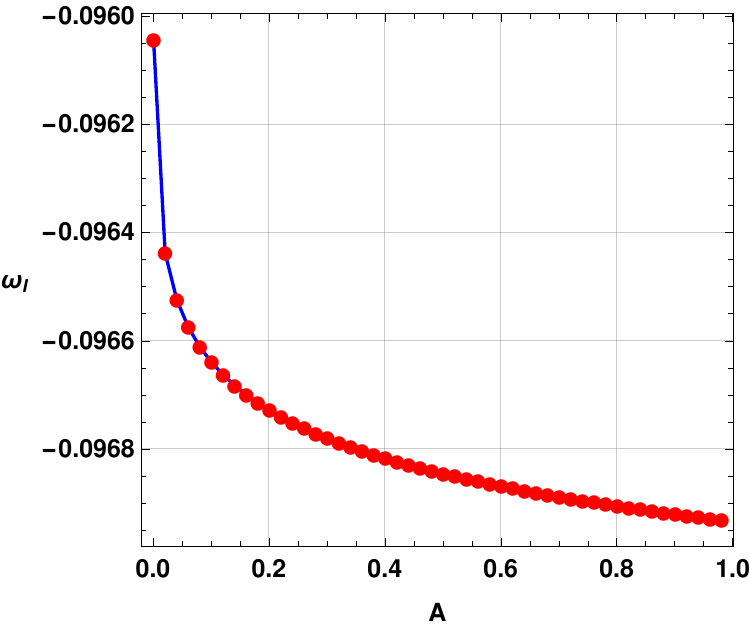} \vspace{-0.2cm}
   \end{center}
\caption{Variation of quasinormal modes with respect to model parameter $A$
with $M=1$, $n= 0$, $\eta = 100$ and $l=3$ for scalar
perturbation.}
\label{QNMs01s}
\end{figure*}

\begin{figure*}[htbp!]
\begin{center}
   \includegraphics[scale = 0.50]{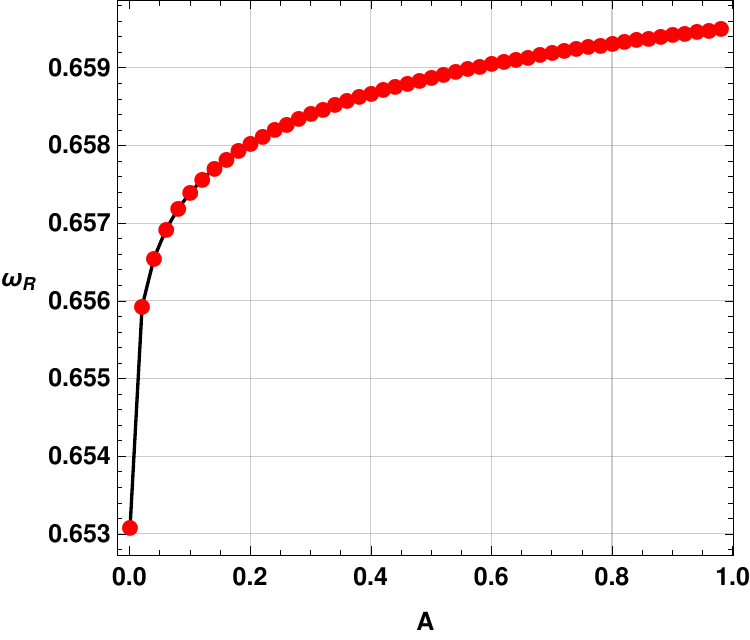}\hspace{0.5cm}
   \includegraphics[scale = 0.50]{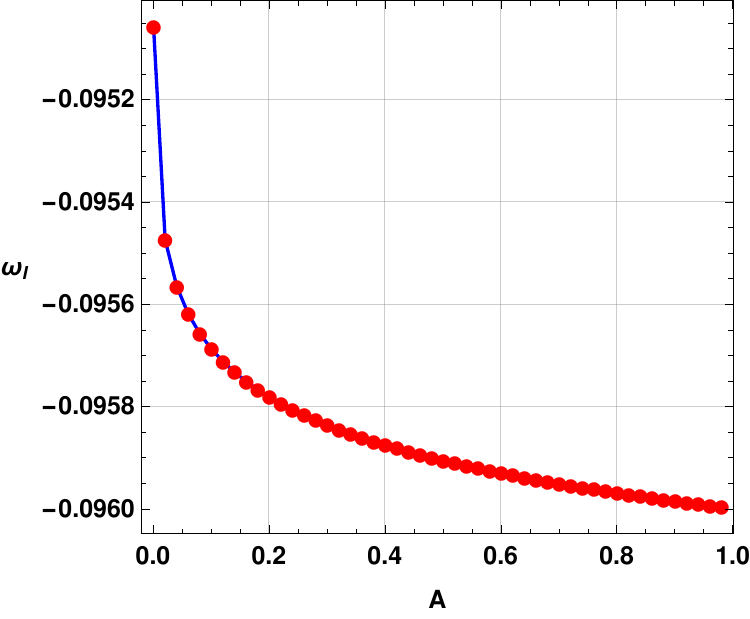} \vspace{-0.2cm}
   \end{center}
\caption{Variation of quasinormal modes with respect to model parameter $A$
with $M=1$, $n= 0$, $\eta = 100$ and $l=3$ for electromagnetic
perturbation.}
\label{QNMs01}
\end{figure*}

\begin{figure*}[htbp!]
\begin{center}
   \includegraphics[scale = 0.50]{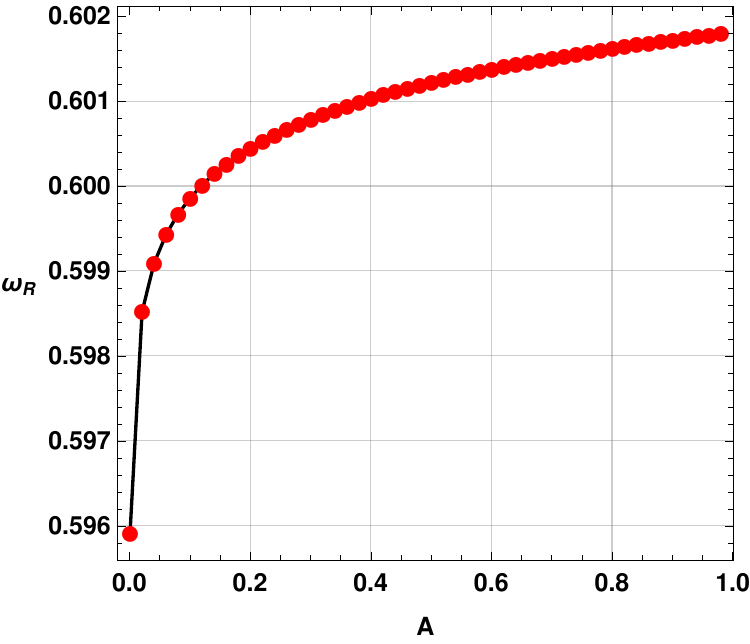}\hspace{0.5cm}
   \includegraphics[scale = 0.50]{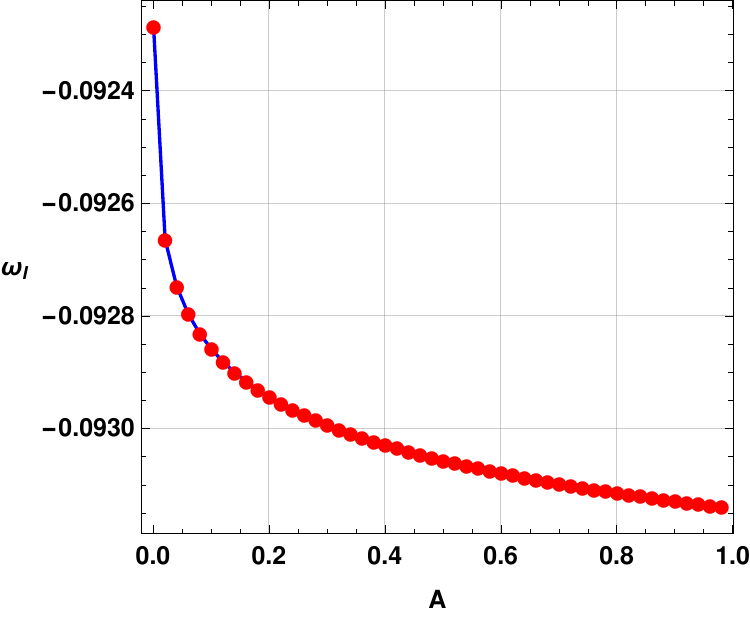} \vspace{-0.2cm}
   \end{center}
\caption{Variation of quasinormal modes with respect to model parameter $A$
with $M=1$, $n= 0$, $\eta = 100$ and $l=3$ for gravitational
perturbation.}
\label{QNMs01g}
\end{figure*}

\begin{figure*}[htbp!]
\begin{center}
   \includegraphics[scale = 0.50]{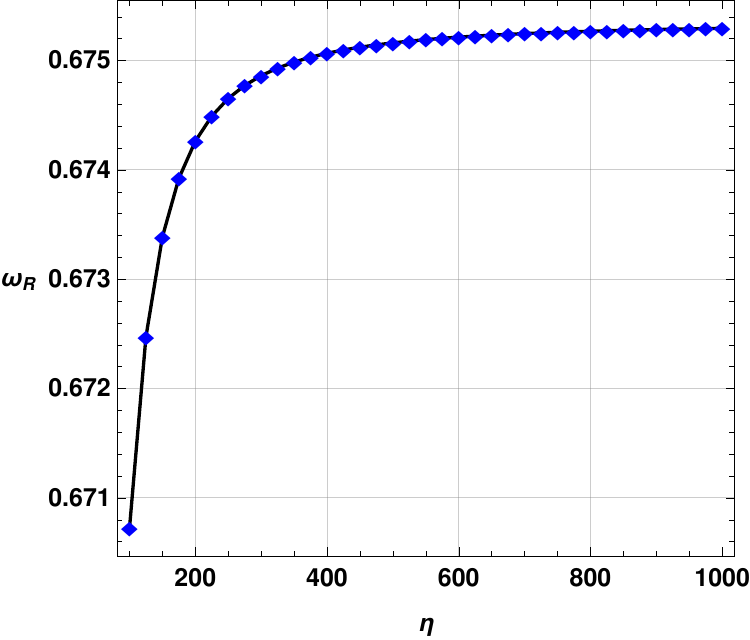}\hspace{0.5cm}
   \includegraphics[scale = 0.50]{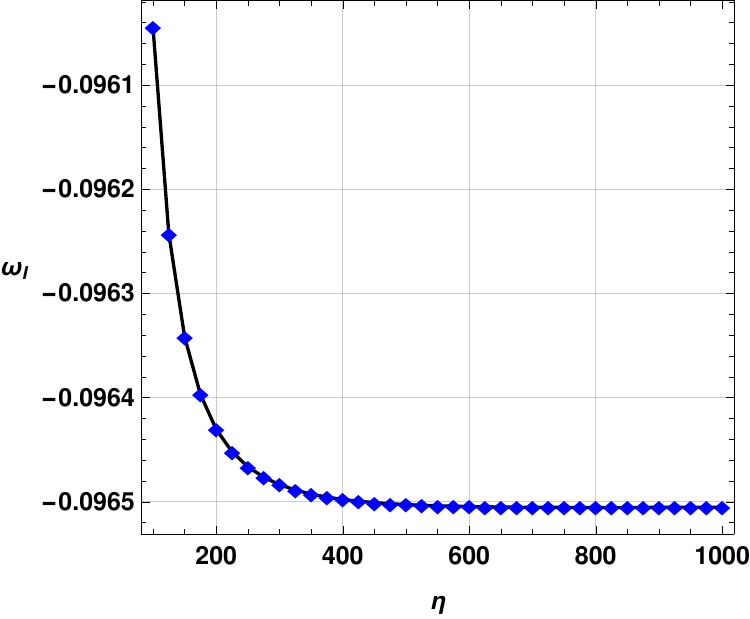} \vspace{-0.2cm}
   \end{center}
\caption{Variation of quasinormal modes with respect to model parameter $\eta$
with $M=1$, $n= 0$, $A = 0.001$ and $l=3$ for scalar
perturbation.}
\label{QNMs02s}
\end{figure*}

\begin{figure*}[htbp!]
\begin{center}
   \includegraphics[scale = 0.50]{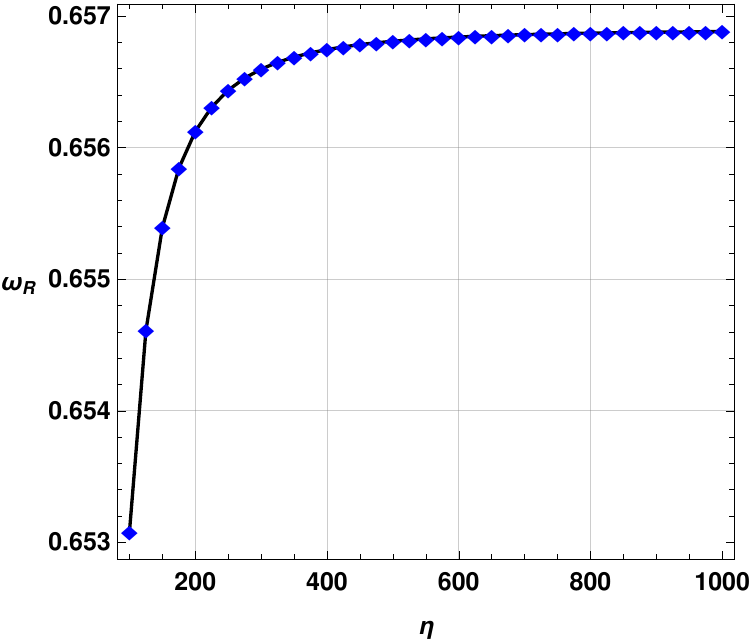}\hspace{0.5cm}
   \includegraphics[scale = 0.50]{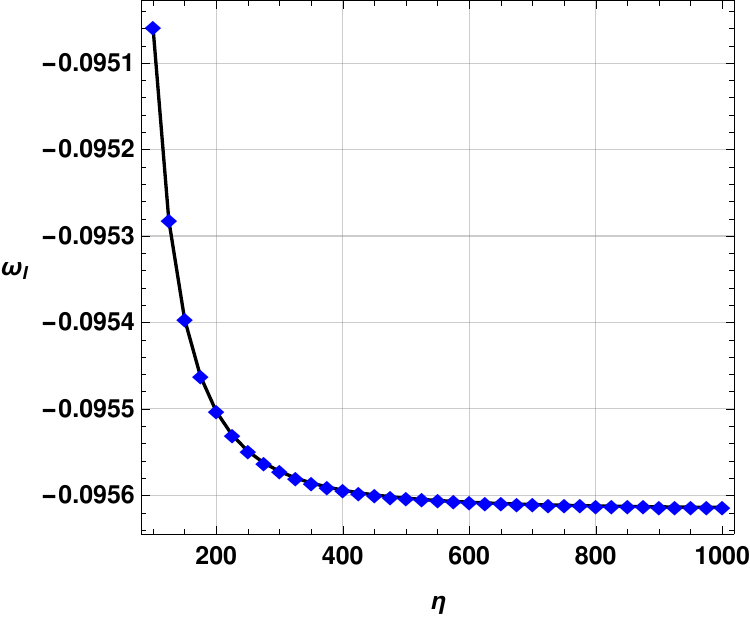} \vspace{-0.2cm}
   \end{center}
\caption{Variation of quasinormal modes with respect to model parameter $\eta$
with $M=1$, $n= 0$, $A = 0.001$ and $l=3$ for electromagnetic
perturbation.}
\label{QNMs02}
\end{figure*}

\begin{figure*}[htbp!]
\begin{center}
   \includegraphics[scale = 0.50]{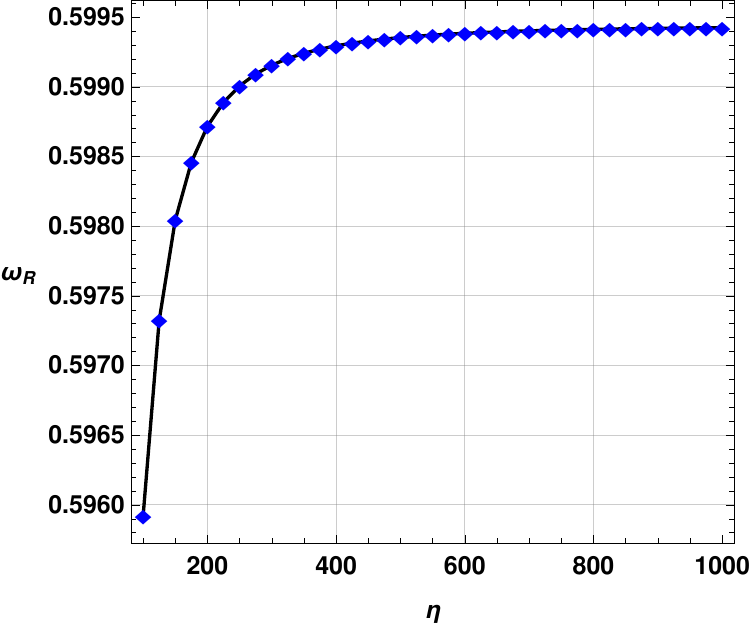}\hspace{0.5cm}
   \includegraphics[scale = 0.50]{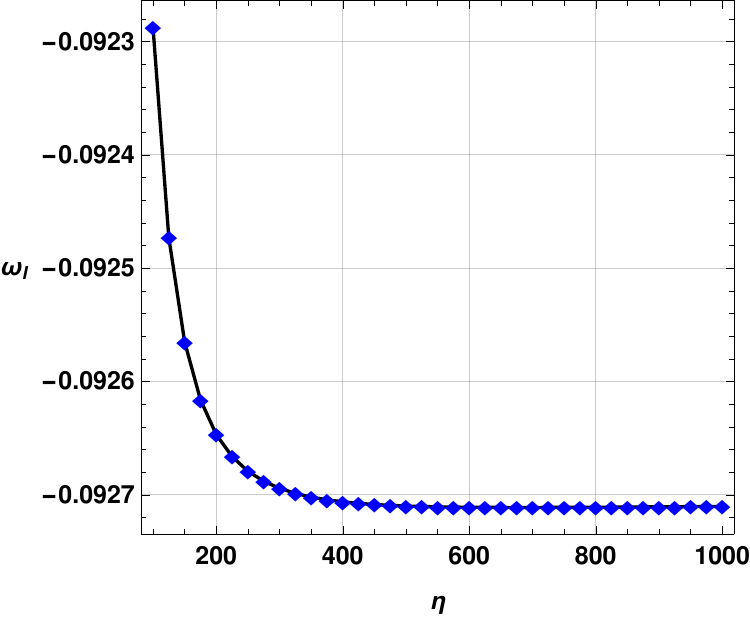} \vspace{-0.2cm}
   \end{center}
\caption{Variation of quasinormal modes with respect to model parameter $\eta$
with $M=1$, $n= 0$, $A = 0.001$ and $l=3$ for gravitational
perturbation.}
\label{QNMs02g}
\end{figure*}

\begin{figure*}[htbp!]
\begin{center}
   \includegraphics[scale = 0.50]{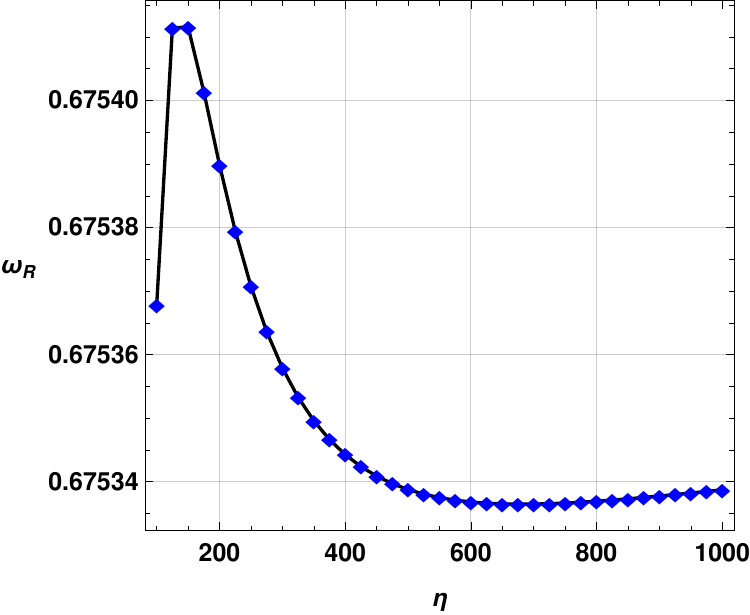}\hspace{0.5cm}
   \includegraphics[scale = 0.50]{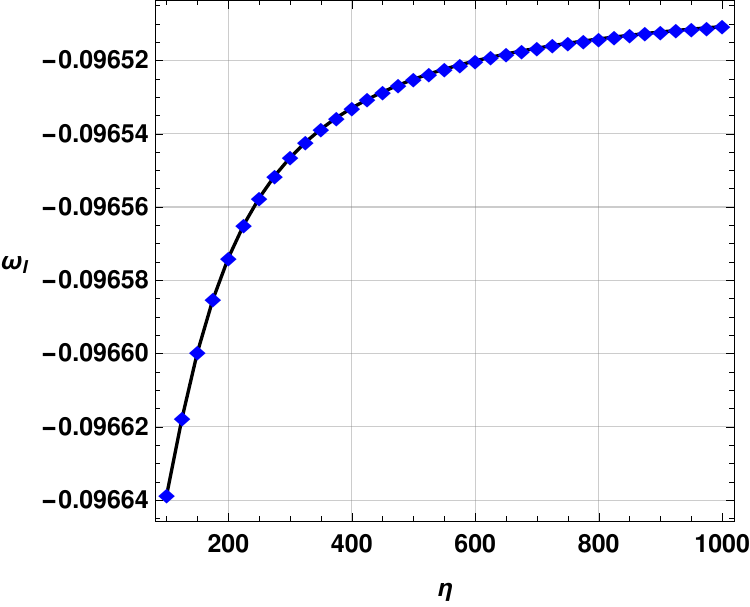} \vspace{-0.2cm}
   \end{center}
\caption{Variation of quasinormal modes with respect to model parameter $\eta$
with $M=1$, $n= 0$, $A = 0.1$ and $l=3$ for scalar
perturbation.}
\label{QNMs02Bs}
\end{figure*}

\begin{figure*}[htbp!]
\begin{center}
   \includegraphics[scale = 0.50]{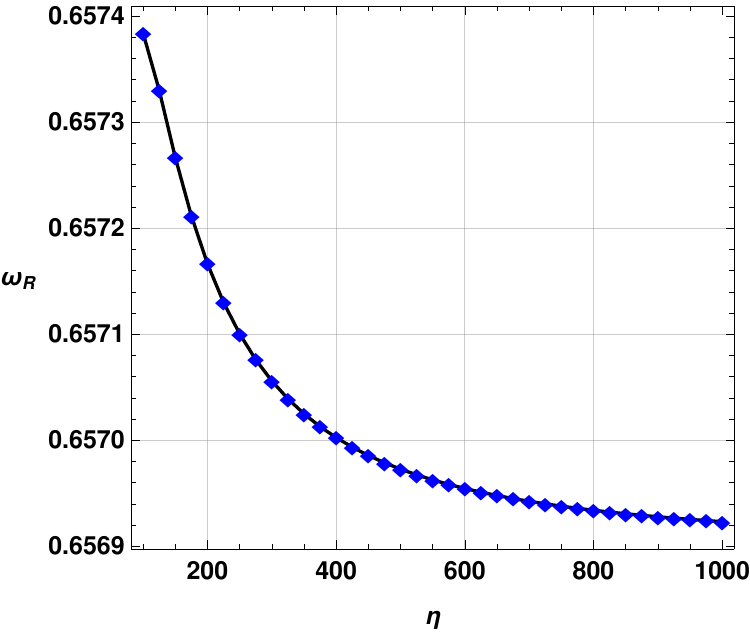}\hspace{0.5cm}
   \includegraphics[scale = 0.50]{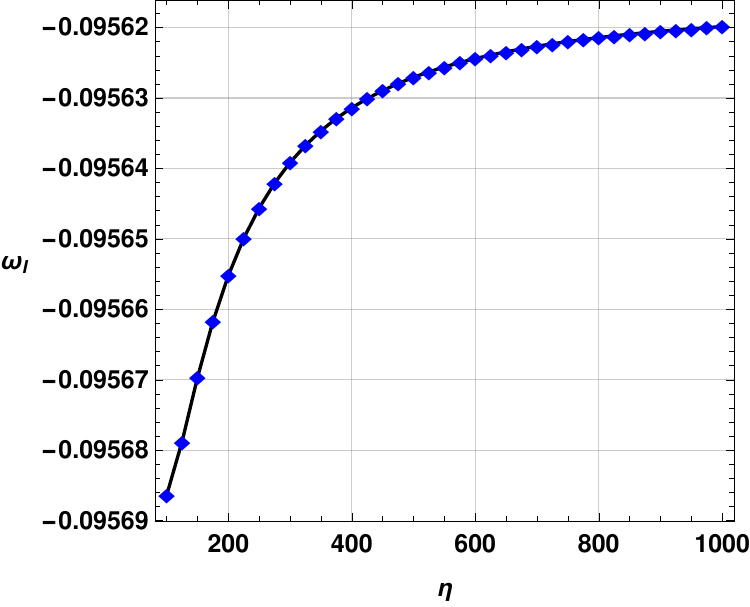} \vspace{-0.2cm}
   \end{center}
\caption{Variation of quasinormal modes with respect to model parameter $\eta$
with $M=1$, $n= 0$, $A = 0.1$ and $l=3$ for electromagnetic
perturbation.}
\label{QNMs02B}
\end{figure*}

\begin{figure*}[htbp!]
\begin{center}
   \includegraphics[scale = 0.50]{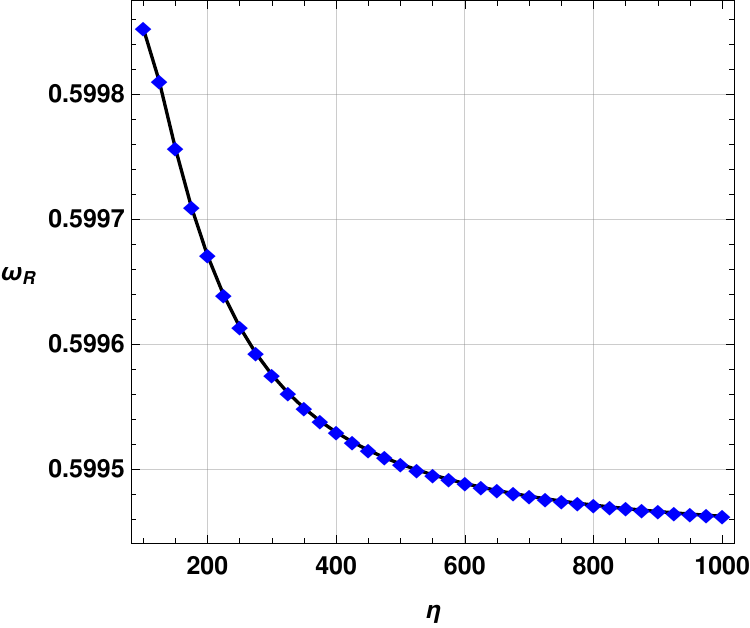}\hspace{0.5cm}
   \includegraphics[scale = 0.50]{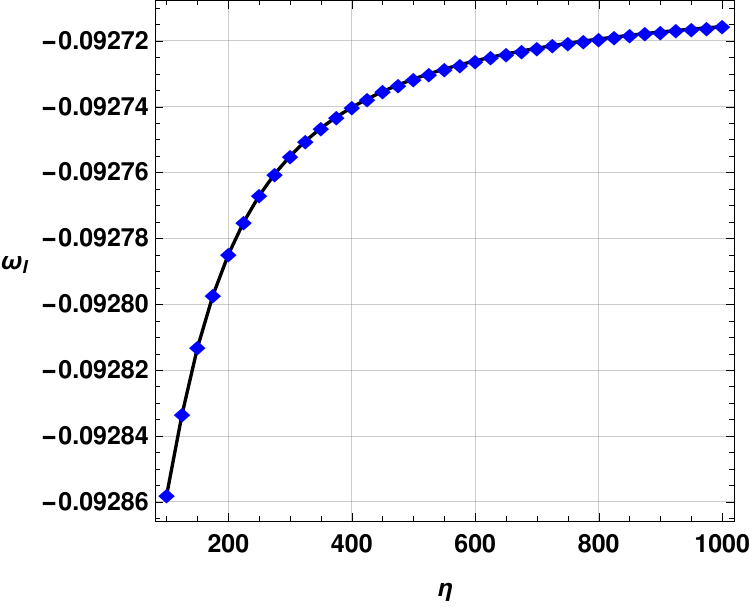} \vspace{-0.2cm}
   \end{center}
\caption{Variation of quasinormal modes with respect to model parameter $\eta$
with $M=1$, $n= 0$, $A = 0.1$ and $l=3$ for gravitational
perturbation.}
\label{QNMs02Bg}
\end{figure*}

In the next stage of our analysis, we examine how the QNM spectrum is influenced by the variation of the model parameters $A$ and $\eta$, which characterise deviations from standard general relativity in the black hole spacetime under consideration. Using the sixth-order WKB approximation enhanced with Padé averaging, we compute the fundamental ($n = 0$) QNMs for fixed values of the mass parameter $M = 1$ and multipole number $l = 3$. The results are shown in Figs. \ref{QNMs01s},\ref{QNMs01},\ref{QNMs01g}, \ref{QNMs02s}, \ref{QNMs02}, \ref{QNMs02g}, \ref{QNMs02Bs}, \ref{QNMs02B} and \ref{QNMs02Bg}, where the real part $\omega_R$ and imaginary part $\omega_I$ of the frequency are plotted separately as functions of $A$ and $\eta$.

Fig. \ref{QNMs01s} illustrates the behaviour of scalar QNMs with respect to the model parameter $A$, which appears logarithmically in the effective potential and influences the potential's asymptotic behaviour. The left panel shows that the real part $\omega_R$ increases monotonically with $A$, indicating that the oscillation frequency of the perturbation becomes higher as $A$ grows. Physically, this suggests that larger values of $A$ correspond to more tightly bound perturbations with faster oscillatory response. The growth of $\omega_R$ is initially sharp for small values of $A$, and the curve gradually saturates at higher $A$, hinting at the onset of an asymptotic behaviour.

On the other hand, the right panel of Fig. \ref{QNMs01s} shows a corresponding decrease in the magnitude of the imaginary part $\omega_I$ with increasing $A$, implying that the damping rate of the perturbation slows down. Since $\omega_I < 0$, a smaller absolute value of $\omega_I$ corresponds to longer-lived modes. Hence, higher values of $A$ not only lead to faster oscillations but also enhance the stability of the electromagnetic perturbations by reducing the decay rate. This reflects a general physical trend: as the effective potential barrier becomes higher or steeper due to changes in $A$, the perturbation becomes more confined, leading to slower energy dissipation.

Figs. \ref{QNMs01} and \ref{QNMs01g} suggest that the impact of $A$ on the QNMs spectrum for electromagnetic and {axial} gravitational perturbations is similar to that of scalar perturbation.

Fig. \ref{QNMs02s} with $A=0.001$, investigates the effect of the model parameter $\eta$ on the scalar QNMs, which governs the strength of the logarithmic correction in the potential and appears in the asymptotic structure as well. In the left panel, we observe that $\omega_R$ increases with $\eta$, though the growth rate becomes slower at large $\eta$. This again indicates that the real part of the frequency is sensitive to the sharpness of the potential barrier. Physically, increasing $\eta$ leads to stronger gravitational backreaction from the modified gravity terms, which shifts the barrier location and affects the wave dynamics. The asymptotic behavior suggests a saturation point beyond which the influence of $\eta$ on the oscillation frequency becomes negligible.

The right panel of Fig. \ref{QNMs02s} shows a rapid decrease in the imaginary part $\omega_I$ with increasing $\eta$, following a sharp fall-off in the small-$\eta$ regime and leveling off in the large-$\eta$ regime. This implies that black holes in high-$\eta$ environments (stronger deviation from general relativity) emit perturbations that decay more rapidly, i.e., the modes become short-lived. This may have observational consequences: in gravitational wave astronomy, such short-lived signals are less likely to be detectable and may be observationally challenging to notice the distinctive imprints of the underlying modified gravity theory.

For electromagnetic and {axial} gravitational perturbations, we observe a similar trend (see Figs. \ref{QNMs02} and \ref{QNMs02g}).

However,  Figs. \ref{QNMs02Bs}, \ref{QNMs02B} and \ref{QNMs02Bg} in which we have considered $A=0.1$, show an opposite picture. In the presence of higher $A$ values, the oscillation frequencies of gravitational waves decrease with an increase in the value of $\eta$, and the damping rate also decreases non-linearly. In such a case, the modes become long-lived and are more likely to be detected. One may note that in the case of scalar perturbation shown in Fig. \ref{QNMs02Bs}, for smaller values of $\eta$, the oscillation frequency of GWs increases rapidly with $\eta$, and after reaching a certain value of $\eta$, the oscillation frequency starts to decrease nonlinearly. Interestingly, for $\eta>700$, it again shows a weak increasing trend.

Overall, the combined analysis of these figures reveals that both the real and imaginary parts of QNMs are strongly dependent on the model parameters. These trends indicate a stabilisation of the system as the deviations from general relativity grow, which could be interpreted as a hallmark signature of the modified gravity framework. The smooth and monotonic behaviour of the QNMs with respect to these parameters also strengthens the theoretical consistency of the model and supports the feasibility of using QNMs as a diagnostic probe for such extended theories of gravity.


\begin{figure*}[htbp!]
\begin{center}
   \includegraphics[scale = 0.68]{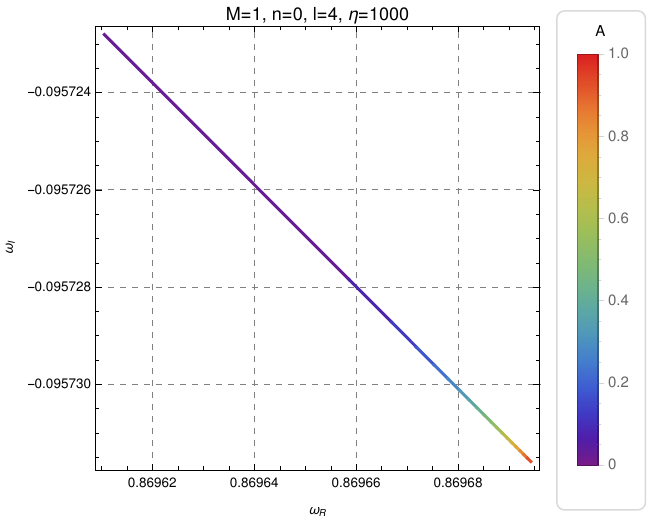}\hspace{0.5cm}
   \includegraphics[scale = 0.68]{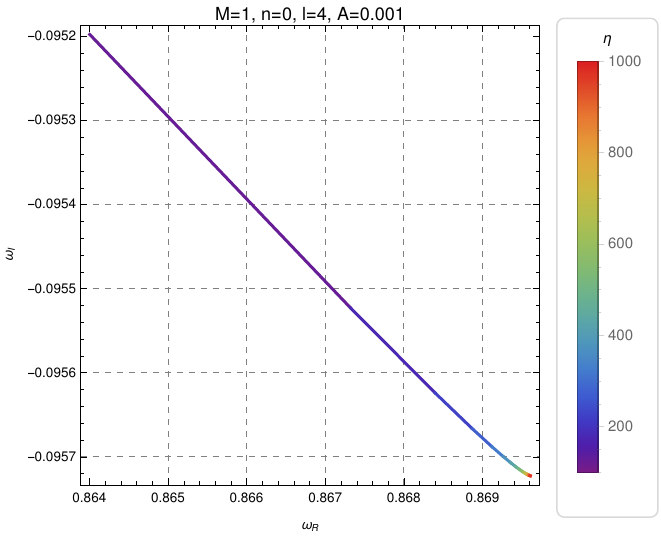} 
   \vspace{-0.2cm}
   \end{center}
\caption{$\omega_I$ vs. $\omega_R$ for scalar QNMs using the Mashhoon method.}
\label{QNMs03Sc}
\end{figure*}

\begin{figure*}[htbp!]
\begin{center}
   \includegraphics[scale = 0.68]{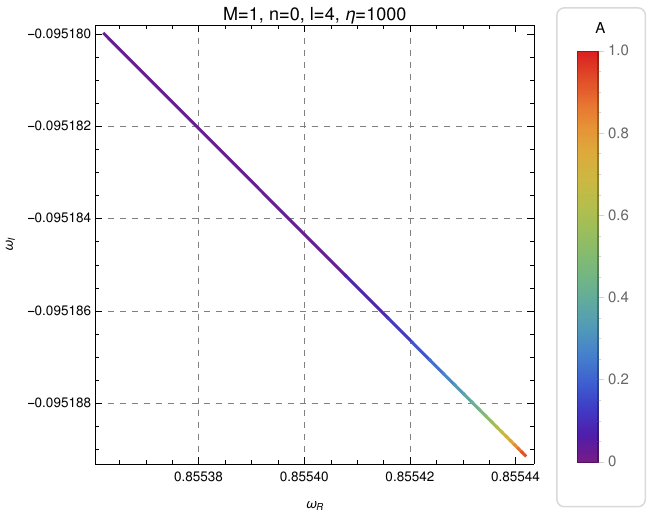}\hspace{0.5cm}
   \includegraphics[scale = 0.68]{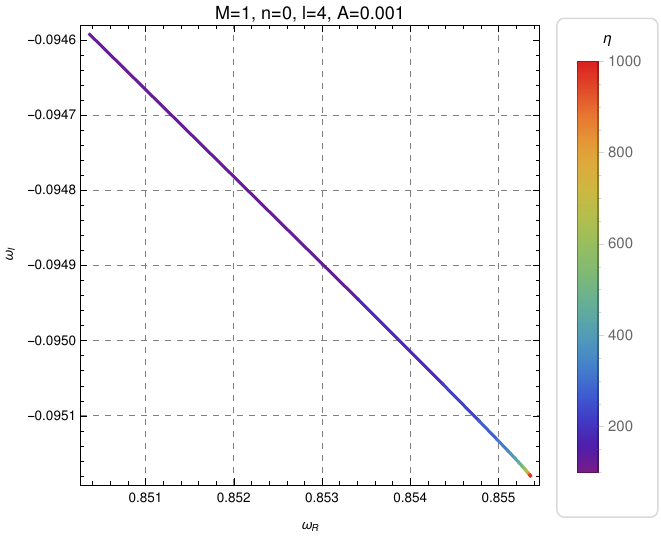} 
   \vspace{-0.2cm}
   \end{center}
\caption{$\omega_I$ vs. $\omega_R$ for electromagnetic QNMs using the Mashhoon method.}
\label{QNMs03}
\end{figure*}

\begin{figure*}[htbp!]
\begin{center}
   \includegraphics[scale = 0.68]{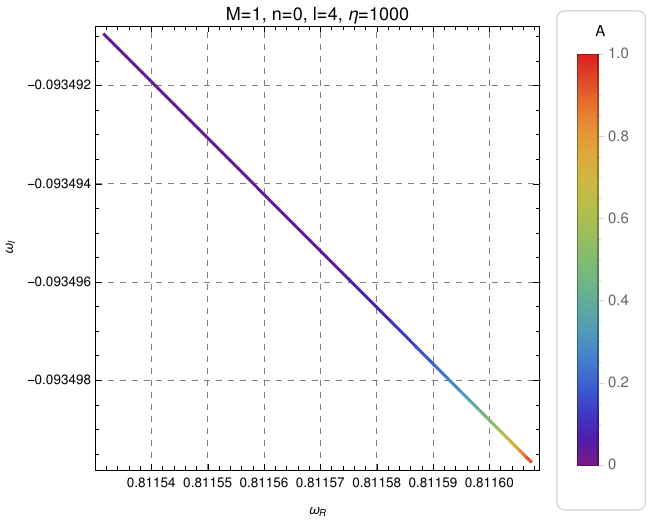}\hspace{0.5cm}
   \includegraphics[scale = 0.68]{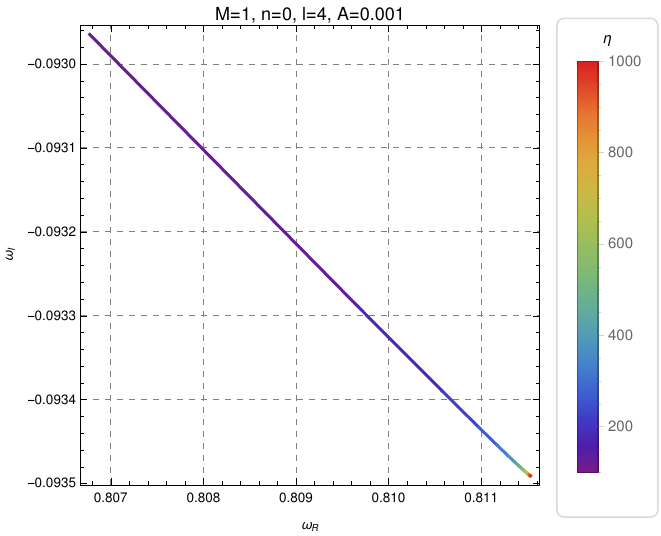} 
   \vspace{-0.2cm}
   \end{center}
\caption{$\omega_I$ vs. $\omega_R$ for gravitational QNMs using the Mashhoon method.}
\label{QNMs03Gr}
\end{figure*}

In Figs. \ref{QNMs03Sc}, \ref{QNMs03} and \ref{QNMs03Gr}, we visualize the behaviour of QNMs in the complex frequency plane by plotting their trajectories under variation of the model parameters $A$ (left panel) and $\eta$ (right panel), using the Mashhoon method. The background spacetime is characterised by fixed values $M = 1$, $n = 0$, and $l = 4$. Each curve in the complex plane represents how the QNM frequency $\omega = \omega_R + i \omega_I$ shifts in response to systematic changes in a single model parameter, while all other parameters are held constant.

The left panels of these Figures correspond to variations in the parameter $A$, with $\eta = 1000$ fixed. As $A$ increases from very small values toward unity, the trajectory of the QNM frequency shifts gradually along a nearly straight line in the complex plane. The real part $\omega_R$ increases monotonically, while the imaginary part $\omega_I$ becomes less negative. This behaviour is consistent with the earlier results presented in Figs. \ref{QNMs01s}, \ref{QNMs01} and \ref{QNMs01g}: increasing $A$ raises the oscillation frequency and decreases the damping rate. The colour gradient illustrates this progression clearly, with darker hues (low $A$) corresponding to lower $\omega_R$ and larger (in magnitude) damping, and lighter hues (high $A$) corresponding to higher $\omega_R$ and smaller damping rates.

The right panels explore the impact of the parameter $\eta$, holding $A = 0.001$ fixed. The behaviour of the QNMs as $\eta$ increases again traces out a smooth and continuous path in the complex plane, mirroring the structure of the left panel. Similar to the case of $A$, increasing $\eta$ causes the real part of the frequency to increase and the imaginary part to move closer to zero, indicating more persistent oscillations. 

This observation aligns with the parameter trends seen in Figs. \ref{QNMs02s}, \ref{QNMs02} and \ref{QNMs02g}. The colour-coded trajectory confirms the strong and monotonic dependence of both $\omega_R$ and $\omega_I$ on the model parameter $\eta$, reinforcing the interpretation that larger $\eta$ stabilises the perturbations and enhances their lifetimes.

These complex-plane contours provide a compact and unified representation of how the electromagnetic QNM frequencies respond to parameter variations. Importantly, they also reinforce the robustness of the Mashhoon approximation, which, despite its simplicity, is capable of capturing the essential features of the QNM evolution across a broad range of modified gravity parameters. The nearly linear paths in both panels also suggest that the real and imaginary components of the QNMs scale coherently with the model parameters, possibly pointing to deeper underlying symmetries or structural properties of the perturbation equations.

\begin{figure*}[htbp!]
   \includegraphics[scale = 0.50]{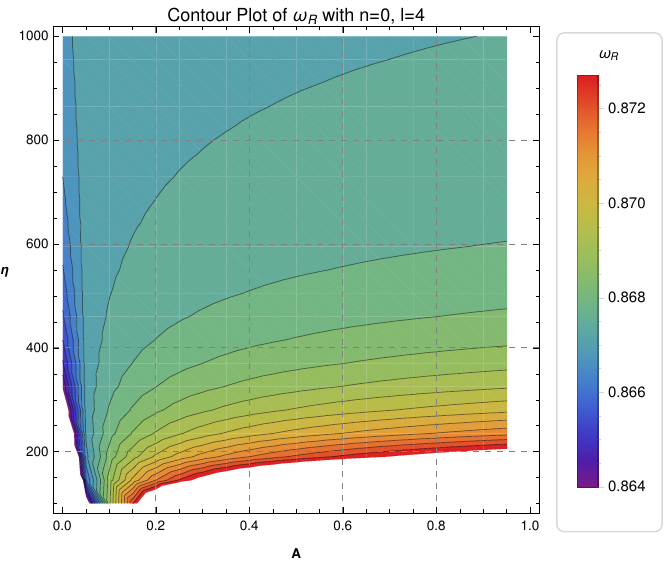}\hspace{0.5cm}
   \includegraphics[scale = 0.50]{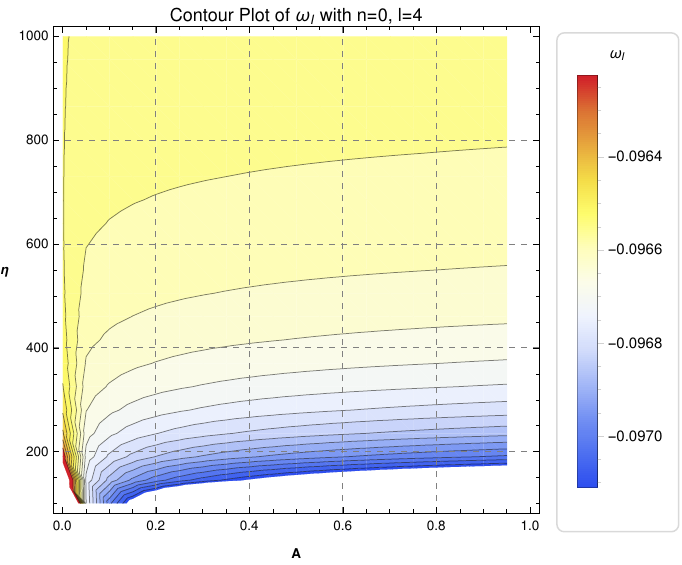}
   \includegraphics[scale = 0.50]{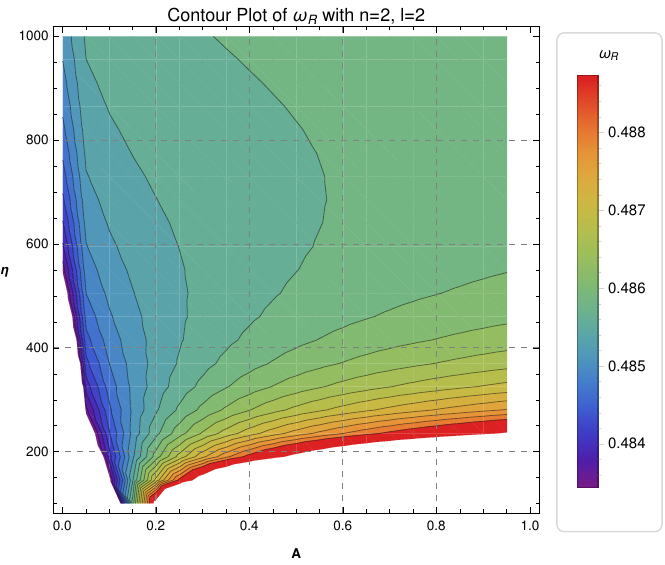}\hspace{0.5cm}
   \includegraphics[scale = 0.50]{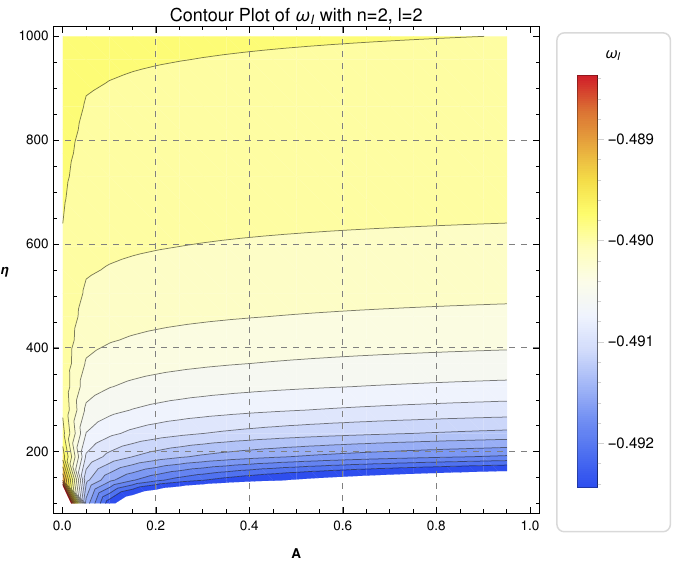}
    \includegraphics[scale = 0.50]{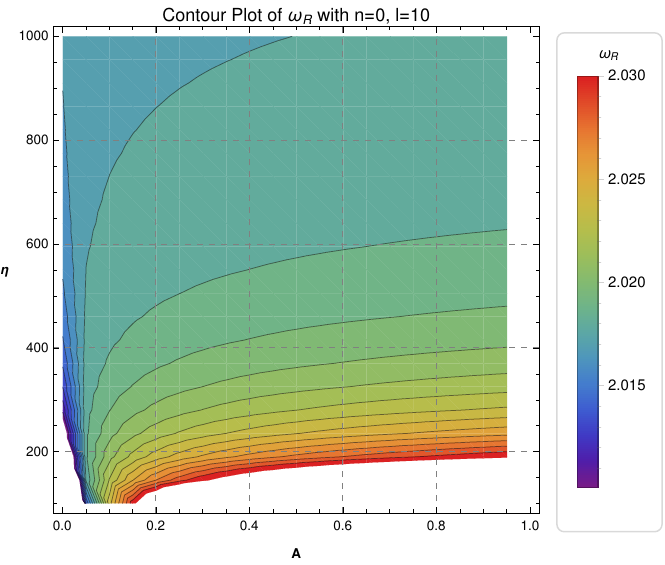}\hspace{0.5cm}
   \includegraphics[scale = 0.50]{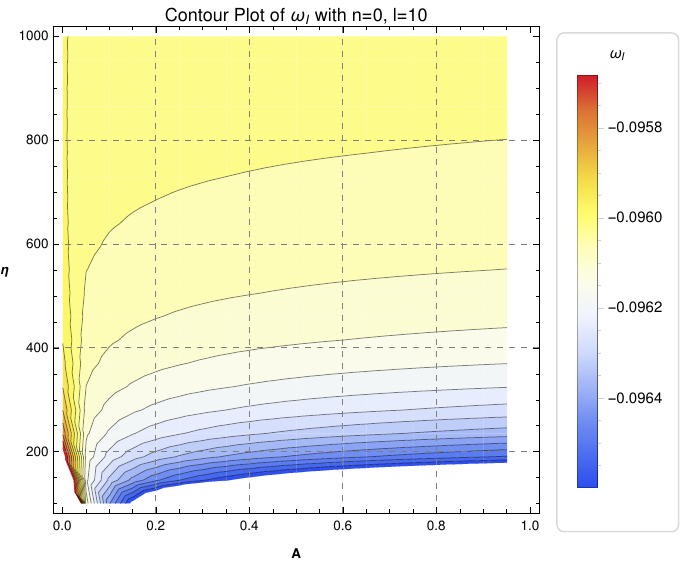}
   \vspace{-0.2cm}
\caption{Contours of scalar QNMs in parameter plane using Mashhoon method.}
\label{QNMs04s}
\end{figure*}

\begin{figure*}[htbp!]
   \includegraphics[scale = 0.50]{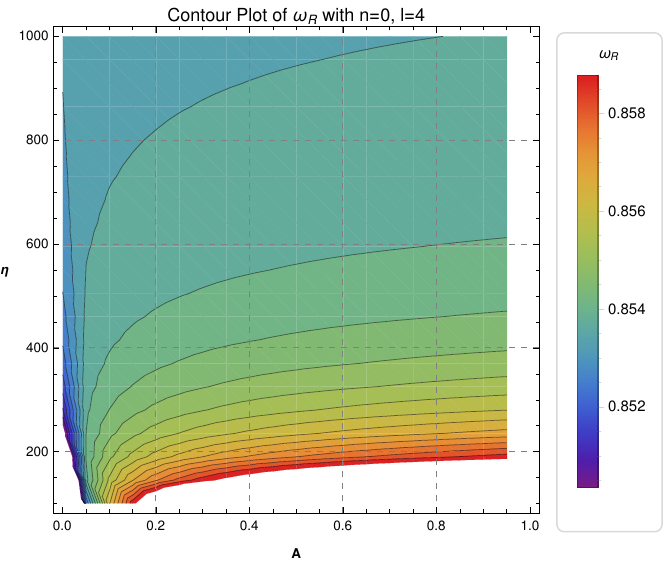}\hspace{0.5cm}
   \includegraphics[scale = 0.50]{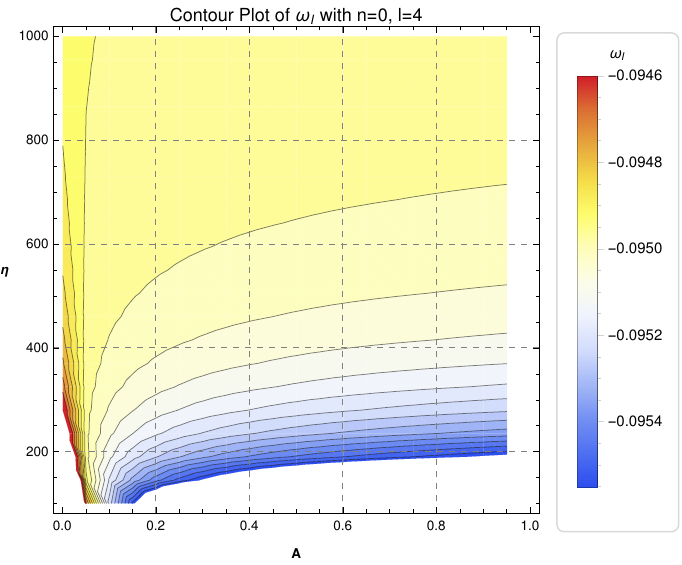}
   \includegraphics[scale = 0.50]{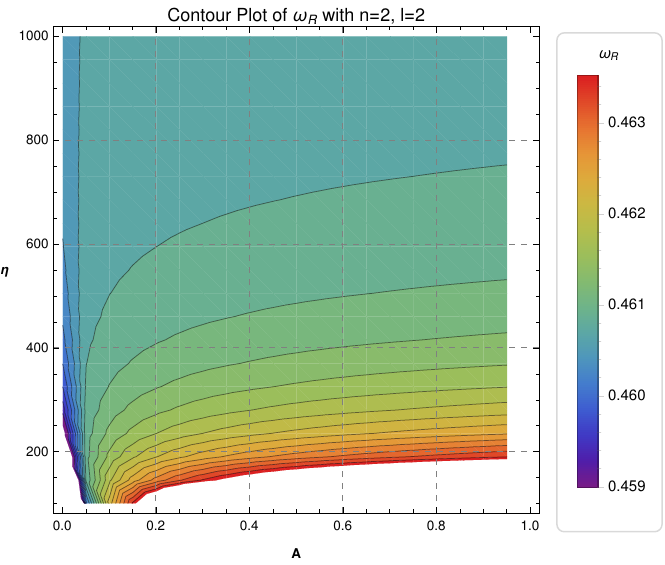}\hspace{0.5cm}
   \includegraphics[scale = 0.50]{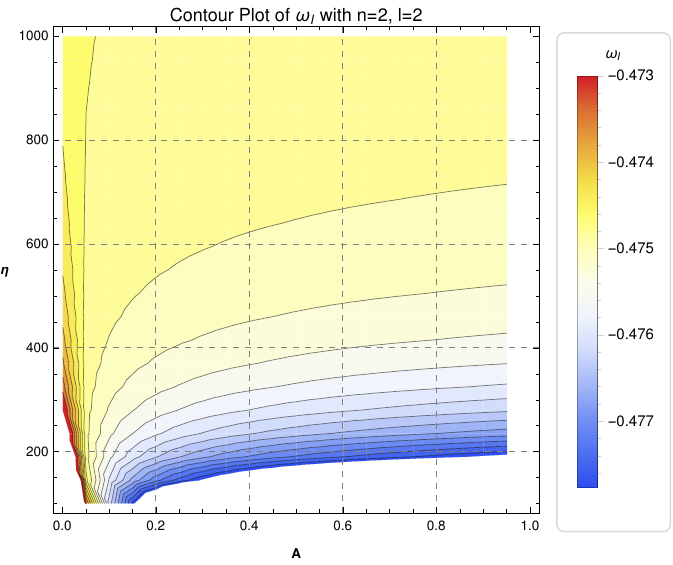}
    \includegraphics[scale = 0.50]{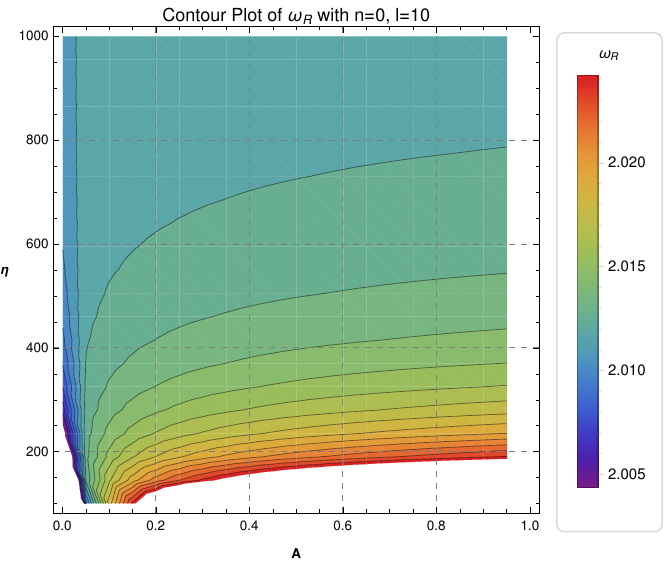}\hspace{0.5cm}
   \includegraphics[scale = 0.50]{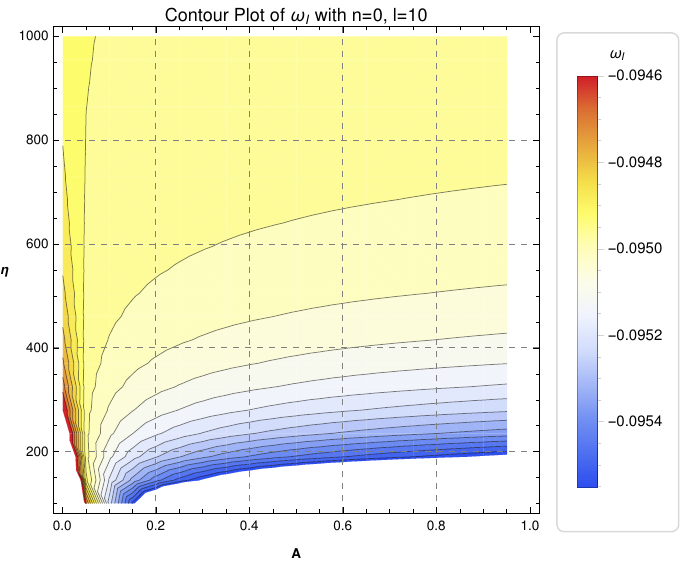}
   \vspace{-0.2cm}
\caption{Contours of electromagnetic QNMs in parameter plane using Mashhoon method.}
\label{QNMs04}
\end{figure*}

\begin{figure*}[htbp!]
   \includegraphics[scale = 0.50]{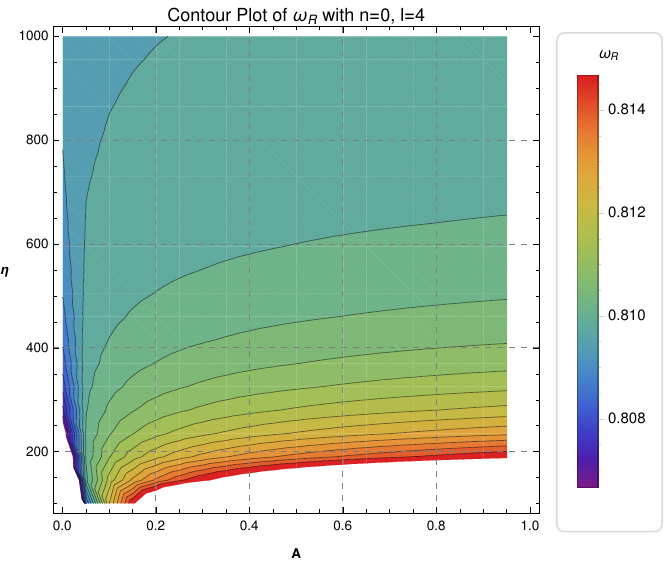}\hspace{0.5cm}
   \includegraphics[scale = 0.50]{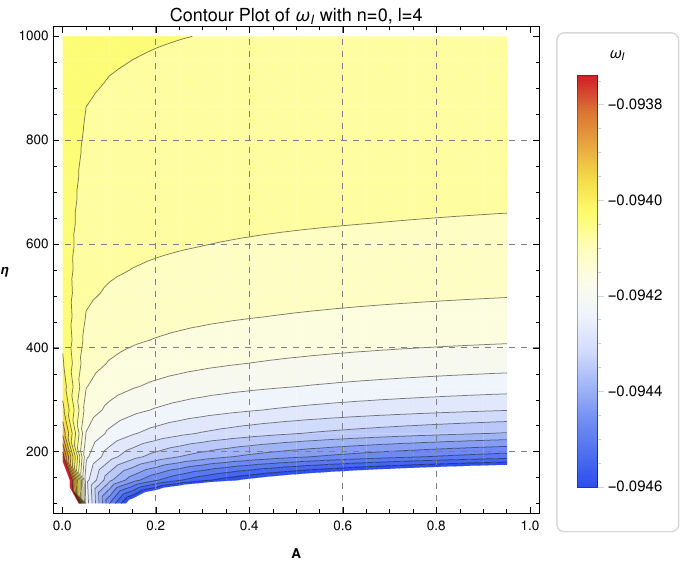}
   \includegraphics[scale = 0.50]{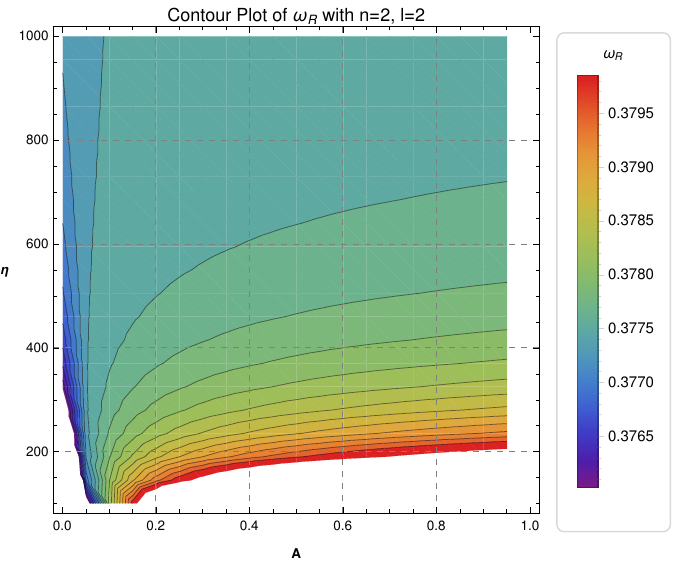}\hspace{0.5cm}
   \includegraphics[scale = 0.50]{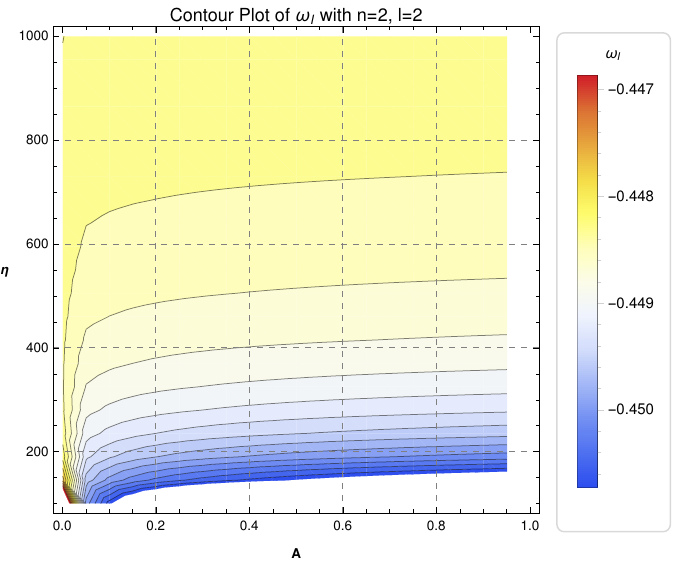}
    \includegraphics[scale = 0.50]{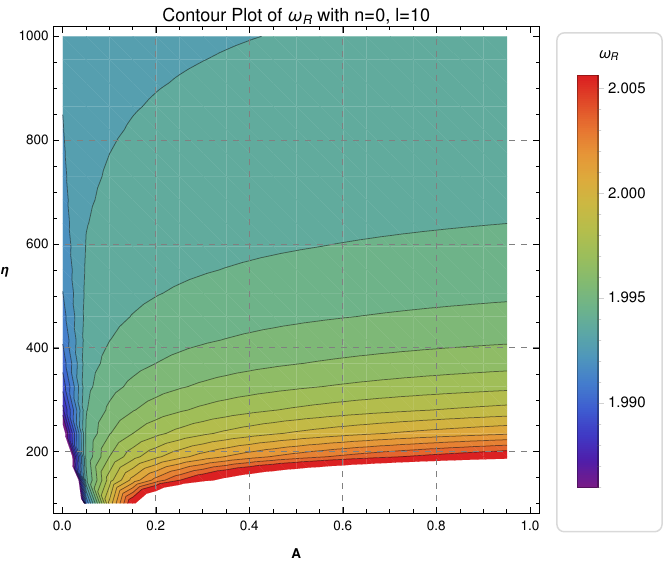}\hspace{0.5cm}
   \includegraphics[scale = 0.50]{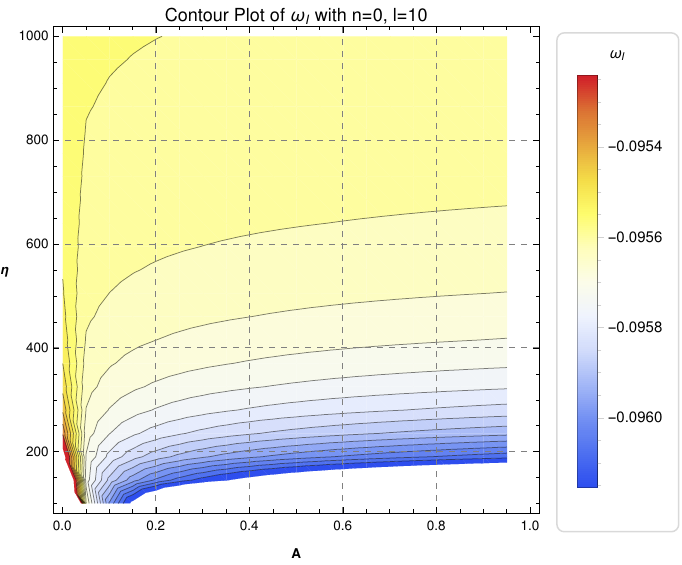}
   \vspace{-0.2cm}
\caption{Contours of gravitational QNMs in parameter plane using Mashhoon method.}
\label{QNMs04g}
\end{figure*}

\subsection{Contour plots}

Figs. \ref{QNMs04s}, \ref{QNMs04} and \ref{QNMs04g} presents the contour plots of the real and imaginary parts of the QNM frequencies in the parameter space of $(A, \eta)$, computed using the Mashhoon method for different values of the overtone numbers and multipole moments. The right panels show that the damping rate $|\omega_I|$ is lowest for small values of both $A$ and $\eta$, and it gradually increases as either parameter increases, indicating that the perturbations become short-lived in strong modified gravity regimes. 

One may note that for higher $A$ values, the damping rate decreases with an increase in the value of $\eta$, and it stands in agreement with the result found in Figs. \ref{QNMs02Bs}, \ref{QNMs02B} and \ref{QNMs02Bg} using the WKB method. The left panels demonstrate that the real part $\omega_R$, associated with the oscillation frequency, increases with an increase in $A$, though the rate of increase diminishes at higher values, highlighting an asymptotic saturation effect. Usually, with an increase in the value of $\eta$, results in a decrease in the oscillation frequency of gravitational waves; however, when $A$ is sufficiently small, an opposite trend is observed. 

This behaviour is consistent with the parameter trends observed in previous results, where monotonic and saturating dependence of QNM frequencies on both $A$ and $\eta$ was established, and further reinforced by the smooth trajectories in the complex plane. Additionally, the variations in the contours for different $n$ and $l$ values suggest how the overtone number and multipole moment influence the QNMs for different perturbations.

\section{Evolution of perturbations on the $f(R)$ black hole geometry}\label{sec5} 

In the previous Section, we have numerically calculated the quasinormal modes and studied their behaviour with respect to the model parameters using 6th order Pad\'e averaged WKB approximation method and the Mashhoon method. In this section, we shall deal with the time domain profiles of the three perturbations. The time domain integration scheme was first introduced in Ref. \cite{gundlach}. To obtain the time evolution profiles in this study, we shall implement the time domain integration formalism following the Ref. \cite{Zhu:2014sya}. For this purpose, we define $\psi(r_*,t) = \psi(i \Delta r_*, j \Delta t) = \psi_{i,j} $ and $V(r(r_*)) = V(r_*,t) =
V_{i,j}$. Now, one can express the perturbation equation associated with the vector potential as
\bea
\dfrac{\psi_{i+1,j} - 2\psi_{i,j} + \psi_{i-1,j}}{\Delta r_*^2} 
&-& \dfrac{%
\psi_{i,j+1} - 2\psi_{i,j} + \psi_{i,j-1}}{\Delta t^2}\nonumber\\
&& - V_i\psi_{i,j} = 0.
\eea

We set the initial conditions as 
\be
\psi(r_*,t) = \exp \left[ -\dfrac{(r_*-k_1)^2%
}{2\sigma^2} \right],
\ee
 and 
 \be
 \psi(r_*,t)\vert_{t<0} = 0, 
 \ee
respectively,  where $k_1$ and $%
\sigma$ are median and width of the initial wave-packet, and then we calculate the time evolution of the scalar field as 
\bea
\psi_{i,j+1} &=& -\,\psi_{i, j-1} + \left( \dfrac{\Delta t}{\Delta r_*}
\right)^2 (\psi_{i+1, j + \psi_{i-1, j}})\nonumber\\
&& + \left( 2-2\left( \dfrac{\Delta t%
}{\Delta r_*} \right)^2 - V_i \Delta t^2 \right) \psi_{i,j}.
\eea

Using the above iteration scheme and choosing a fixed value of $\frac{\Delta t}{\Delta r_*}$, one can easily obtain the profile of $\psi$ with respect to time $t$. However, one should keep $\frac{\Delta t}{\Delta r_*}
< 1$ so that the Von Neumann stability condition is satisfied during the numerical procedure.

\begin{figure*}[htbp!]
   \includegraphics[scale = 0.62]{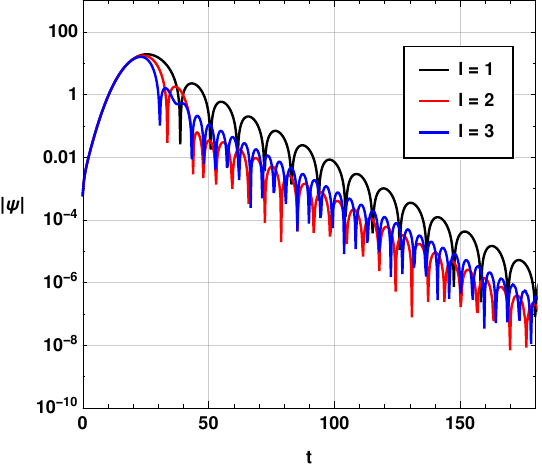}
   \includegraphics[scale = 0.62]{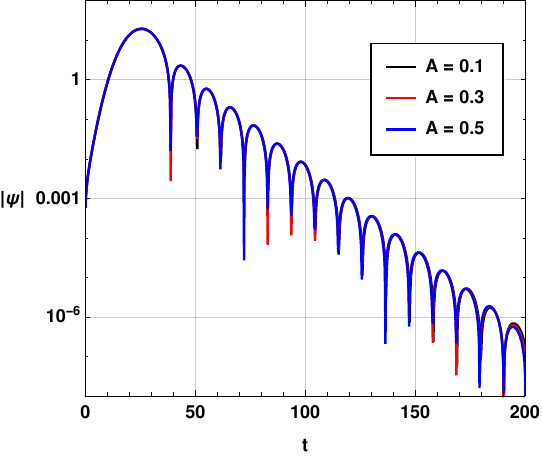}
    \includegraphics[scale = 0.62]{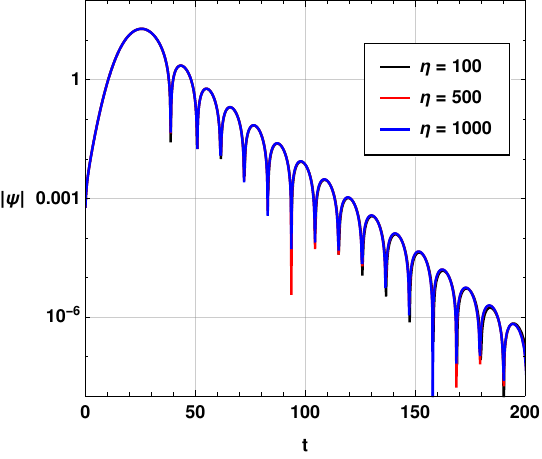}
\caption{Time domain for fundamental scalar
perturbations with $M=1$. On the first panel, $A=0.1, \eta = 100.$ On the second panel, $l=1, \eta = 100$. On the third panel $l=1, A=0.1$.}
\label{TD01}
\end{figure*}

\begin{figure*}[htbp!]
   \includegraphics[scale = 0.62]{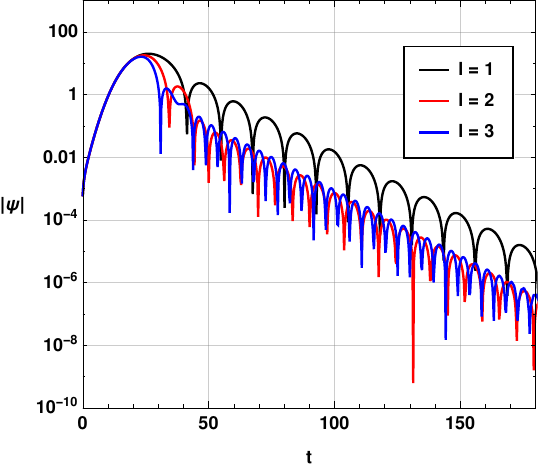}
   \includegraphics[scale = 0.62]{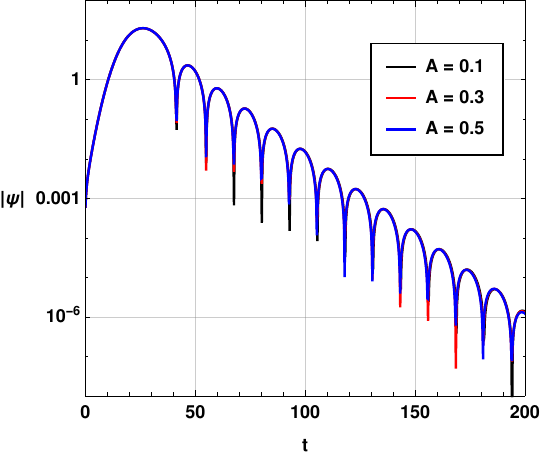}
    \includegraphics[scale = 0.62]{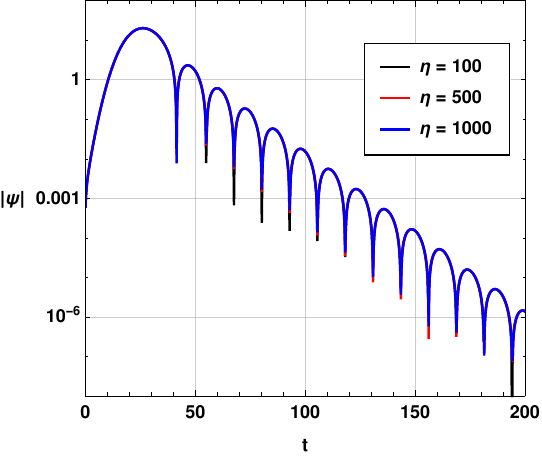}
\caption{Time domain for fundamental electromagnetic
perturbations with $M=1$. On the first panel, $A=0.1, \eta = 100.$ On the second panel, $l=1, \eta = 100$. On the third panel $l=1, A=0.1$.}
\label{TD02}
\end{figure*}

\begin{figure*}[htbp!]
   \includegraphics[scale = 0.62]{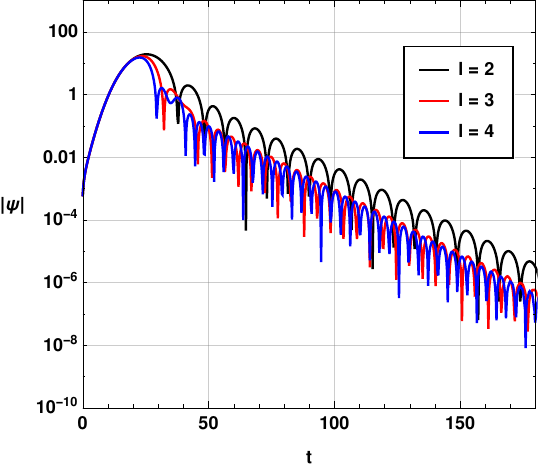}
   \includegraphics[scale = 0.62]{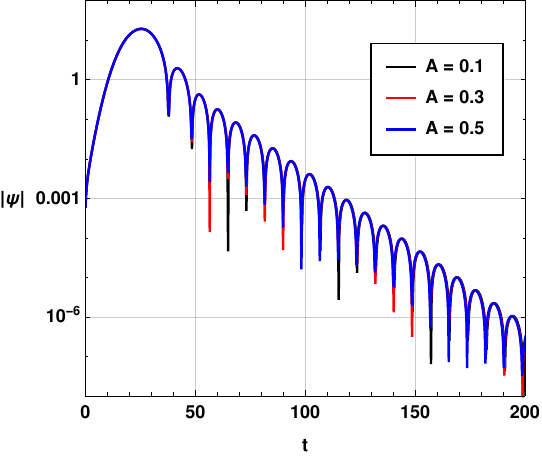}
    \includegraphics[scale = 0.62]{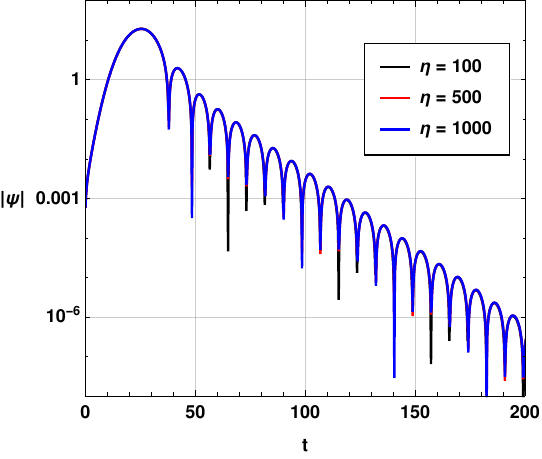}
\caption{Time domain for fundamental gravitational
perturbations with $M=1$. On the first panel, $A=0.1, \eta = 100.$ On the second panel, $l=2, \eta = 100$. On the third panel $l=2, A=0.1$.}
\label{TD03}
\end{figure*}

The time domain profiles associated with the scalar, vector and {axial} gravitational perturbations are shown in Figs. \ref{TD01}, \ref{TD02} and \ref{TD03}. The first panel in each of them shows the impacts of the multipole moment on the QNMs. The second panel shows the impacts of the model parameter $A$ on the evolution of time domain profiles. Finally, the third panel shows the impacts of the parameter $\eta$ on the time domain profiles. As evident from the previous investigation by using the WKB method and the Mashhoon method, the impact of the parameters on the QNMs is very small. These evolution profiles stand in agreement with the results obtained in the previous section.

\section{Concluding Remarks}
\label{sec6}

In this work, we have investigated the scalar, electromagnetic, and gravitational QNMs of the generalized vacuum Schwarzschild–de Sitter black hole solution arising in the context of $f(R)$ gravity. The solution represents a generalization of the standard Schwarzschild-de Sitter solution of general relativity, and it adds two new terms to the general relativistic solution. The solution depends on a set of four arbitrary integration constants. A very interesting property of the solution is that it is not asymptotically flat, even in the absence of the term proportional to $r^2$, which we can interpret as corresponding to a cosmological constant. The perturbation spectra of black hole beyond standard general relativity  could exhibit complex structures with long-lived modes that generate echo-like modulations on the ringdown signal. Moreover, theoretical investigations have shown that the emergence of new QNM families is a general effect in many  physical models. Hence the analysis of the quasinormal mode spectrum can give important insights into the nature of black holes, and of the underlying gravitational theory. 

To compute the QNMs of the generalized Schwarzschild-de Sitter solution of the $f(R)$ gravity theory, we have used the well-established higher-order WKB approximation method, together with an analytical approach we call the Mashhoon method. A systematic comparison with the corresponding QNMs of the Schwarzschild black hole in general relativity revealed that the spectra remain remarkably close across all three types of perturbations. This result highlights the robustness of the Schwarzschild spectrum even under the modifications introduced by $f(R)$ gravity.

We have also compared in detail the QNMs obtained from the WKB and Mashhoon methods. It was found that for lower multipole moments, the difference between the two approaches becomes more pronounced. This can be attributed to two main reasons: (i) the accuracy of the WKB method decreases when $l-n$ is small, leading to larger errors, and (ii) the Mashhoon method inherently carries higher uncertainties in this regime. For larger values of $l$, however, the agreement between the two methods improves significantly, thereby reinforcing their applicability in the semiclassical limit.

In the next phase of our analysis, we have explored the effects of the model parameters $A$ and $\eta$ on the QNMs. Our results clearly demonstrate that both parameters influence the oscillation frequency and the damping rate in a highly nonlinear fashion. Specifically, for small values of $A$, both parameters tend to enhance the oscillation frequency and damping, producing results close to the Schwarzschild case. For larger values of $A$, however, the scalar perturbations exhibit a distinct departure from the electromagnetic and {axial} gravitational perturbations, signaling the presence of unique features associated with the scalar sector in $f(R)$ gravity. These observations were further substantiated through contour plots of the QNMs in the parameter space, which confirmed the consistency of the spectral behavior across different regions of parameter values.

An important outcome of our study is that, throughout all parameter choices and perturbation types, the imaginary parts of the QNMs remained negative. This ensures that the generalized Schwarzschild–de Sitter black hole in $f(R)$ gravity is stable against scalar, electromagnetic, and {axial} gravitational perturbations. To further substantiate this, we also constructed time-domain profiles of the perturbations, which provided direct visualization of the ringdown phase and confirmed the expected decay of gravitational wave signals.

Taken together, our analysis shows that while the QNM spectrum in $f(R)$ gravity closely resembles that of general relativity for certain parameter ranges, subtle nonlinear deviations emerge in others, particularly in the scalar sector. Such deviations may serve as potential observational signatures of modified gravity. With the increasing sensitivity of present and upcoming gravitational wave detectors, these small but distinctive modifications could, in principle, be detected, thereby offering an avenue to test $f(R)$ gravity in the strong-field regime. Moreover, in comparison to other extended theories of gravity, such as scalar–tensor and Gauss–Bonnet models, the QNMs in $f(R)$ gravity tend to remain closer to the Schwarzschild case, suggesting that while their detection may be more challenging, they carry equally significant implications for probing deviations from general relativity.

 While the present work establishes the foundational theoretical spectrum of quasinormal modes for the generalized Schwarzschild-de Sitter black hole in $f(R)$ gravity, we acknowledge that further validation is necessary to transition these theoretical findings into testable predictions. Specifically, rigorous quantitative comparisons with observational data are required to establish empirical constraints on the model parameters $A$ and $\eta$. Future studies employing Bayesian parameter estimation against actual gravitational wave ringdown signals---such as those currently detected by the LIGO-Virgo-KAGRA collaboration, or those anticipated from future observatories like LISA and the Einstein Telescope---will be essential. Such empirical analyses will place stringent bounds on $A$ and $\eta$, ultimately determining the observational viability of this specific $f(R)$ gravity model and its distinct phenomenological signatures in the strong-field regime.

{ While our study provides a comprehensive analysis of the scalar, electromagnetic, and axial gravitational perturbations, we acknowledge that a complete treatment of gravitational perturbations in modified $f(R)$ gravity would ideally encompass the even-parity (polar) sector alongside the extra scalar degree of freedom, often referred to as the scalaron. However, extracting decoupled wave equations for the polar parity and the scalaron in generic $f(R)$ gravity presents significant mathematical complexities. Specifically, for a generalized background metric where the Ricci scalar $R$ is non-constant, the polar perturbations are strongly coupled to the fluctuations of the scalaron field. Decoupling these sectors generally requires assuming a highly specific functional form for $f(R)$, which falls outside the scope of our current approach that focuses on the phenomenological signatures of the exact geometry itself. Therefore, the full integration and analysis of the strongly coupled polar and scalaron modes represent a highly compelling avenue for our future research.}

Future research may focus on extending this analysis to rotating black holes in $f(R)$ gravity, which introduce qualitatively richer dynamics such as superradiance. Investigations in the eikonal limit and connections with photon sphere properties could also provide deeper insights into the geodesic structure of these spacetimes. Furthermore, the inclusion of higher-curvature or quantum corrections might unveil additional fine features in the QNM spectrum beyond the classical level. Ultimately, our results reinforce the idea that QNMs remain one of the most powerful probes of the nature of gravity, with the potential to discriminate between general relativity and modified theories in forthcoming gravitational wave observations.


\end{document}